\documentclass{article}

\usepackage{amsmath, amssymb, amsthm}
\usepackage{graphicx}
\usepackage{enumitem}
\usepackage{geometry}
\usepackage{graphicx}
\usepackage{subcaption}
\usepackage{algorithm}
\usepackage{algorithmic}
\usepackage{amsmath} % for improved math formatting
\usepackage{xcolor}
\usepackage{tabularx}
\usepackage{booktabs}
\usepackage{multirow}
\usepackage{amsmath}
\usepackage{comment}
\usepackage{float}
\DeclareMathOperator*{\argmin}{argmin}

\usepackage[english]{babel}
\usepackage{amsthm}
\newtheorem{theorem}{Theorem}[section]

\newtheorem{assumption}[theorem]{Assumption}
\newtheorem{remark}[theorem]{Remark}

\newcommand{\RR}[0]{\mathbb{R}}

\newcommand{\PP}[0]{\mathbb{P}}
\newcommand{\NN}[0]{\mathcal{N}}

\newcommand{\Var}[0]{\text{Var}}

\begin{document}

\title{Regularized Targeted Maximum Likelihood Estimation in Highly Adaptive Lasso Implied Working Models}
\author{Yi Li$^1$, Sky Qiu$^1$, Zeyi Wang $^2$,Mark van der Laan$^1$\\
Division of Biostatistics, University of California, Berkeley$^1$\\
Oklahoma State University$^2$}
\date{}
\maketitle

\begin{abstract}
We address the challenge of performing Targeted Maximum Likelihood Estimation (TMLE) after an initial Highly Adaptive Lasso (HAL) fit. Existing approaches that utilize the data-adaptive working model selected by HAL—such as the \emph{relaxed HAL} update—can be simple and versatile but may become computationally unstable when the HAL basis expansions introduce collinearity. \emph{Undersmoothed HAL} may fail to solve the efficient influence curve (EIC) at the desired level without overfitting, particularly in complex settings like survival‐curve estimation. A full HAL‑TMLE, which treats HAL as the initial estimator and then targets in the nonparametric or semiparametric model, typically demands costly iterative clever‑covariate calculations in complex set-ups like survival analysis and longitudinal mediation analysis.

To overcome these limitations, we propose two new HAL-TMLEs that operate within the finite‐dimensional working model implied by HAL: \textbf{Delta‑method regHAL-TMLE} and \textbf{Projection‑based regHAL-TMLE}. We conduct extensive simulations to demonstrate the performance of our proposed methods.

\end{abstract}

\vspace{-1em}
\section{Introduction}\label{sec:intro}

Targeted Maximum Likelihood Estimation (TMLE) is a general methodology for constructing asymptotically linear and efficient estimators by updating initial estimates, under the assumption that the second-order remainder - typically involving a cross-product of nuisance parameter estimation errors - is $o_P(n^{-1/2})$. The Highly Adaptive Lasso (HAL) achieves convergence rates of $O_P(n^{-2/3})$ up to $\log n$ factors in loss-based dissimilarity (which generally bounds the second-order remainder), requiring only c\`adl\`ag and finite variation functions. Thus, using HAL as the initial estimator in TMLE (HAL-TMLE) yields a class of consistent and asymptotically linear estimators under mild and realistic conditions. 

Recent developments on adaptive modeling improve bias-variance trade-off by data-adaptively selecting a working model that approximates an oracle submodel simpler than the full nonparametric model. This approach is particularly useful in settings such as estimating treatment effects with limited overlap or incorporating real-world data into randomized trials, where non-adaptive estimators suffer from high variance and instability due to population differences and not adapting to the complexity of true data generating process. For example, relaxed HAL (refitting coefficients with relaxed variation bounds for HAL basis selected under bounded variation) constructs computationally feasible adaptive HAL-TMLE updates with respect to the model spanned by the selected basis in the initial HAL fit. However, relaxed HAL may become computationally unstable when the HAL basis expansions introduce collinearity. 

\paragraph{Why this problem?} %
When a \emph{Highly Adaptive Lasso} (HAL) fit is used as the starting point for
Targeted Maximum Likelihood Estimation (TMLE), two sources of
instability dominate practice:

\begin{enumerate}[itemsep=2pt,label=(\alph*)]
\item \textbf{Collinearity inside the HAL basis.}  
      The Fisher information for the working model can be singular, so a
      HAL–maximum-likelihood estimator (HAL-MLE, a “relaxed HAL” update)
      becomes numerically fragile and may overfit relevant directions
      \cite{carroll1998sandwich,tsiatis2006semiparametric,vanderlaan2021efficient}.
\item \textbf{Expensive and instable clever-covariate construction.}  
      If instead one targets in the \emph{non-parametric} model, in the presence of positivity, the clever covariate estimations can vary a lot and are numerically instable. In addition, the clever
      covariate must be recomputed at every step in many longitudinal or survival
      problems and involve 
      numerical integrations that affect targeting step heavily \cite{rytgaard2021onesteptmletargetingcausespecific,wang2025targeted}.
\end{enumerate}

\paragraph{Our idea in one line.} %
\emph{Stay inside the HAL-induced working model and target \underline{only} the
score directions within the working model that matter for the parameter.} The resulting estimators, called \textit{regularised HAL-TMLE}, are both adaptive and more stable, obtained still through a two-stage plug-in procedure that begins with initial HAL fits, but solves only the relevant score equations in the selected working models, and only to a controlled level of precision on finite samples. 

\paragraph{Two regularised HAL–TMLEs.} %
We study two regularized HAL-TMLEs depending on two ways to approximate the parametric efficient influence curve
(EIC) inside the HAL working model $M_n$ for targeting:
\begin{enumerate}[label=(\roman*),itemsep=2pt]
  \item \textbf{\textit{Delta-method regHAL-TMLE}} – add a small ridge
        $\eta I$ before inverting the empirical information matrix
        \cite{vanderlaan2018targeted} to estimate the parametric EIC.
  \item \textbf{\textit{Projection-based regHAL-TMLE}} – regress any valid
        influence function onto the HAL score space with a lasso penalty
        \cite{vanderlaan2021efficient} to estimate the parametric EIC.
\end{enumerate}
Both updates live completely within $M_n$, guarantee first-order bias removal,
and are numerically stable even when working model size $p\gg n$.

\paragraph{Road-map.} %
Section~\ref{sec:preliminary} recalls just enough HAL and TMLE machinery for later
use. Section~\ref{sec: set up methodology}–\ref{sec: EIC approximation} constructs the two regularised updates as regHAL-TMLE;
Section~\ref{sec:theory} states main theorems (proofs in
Appendix~\ref{sec: proof appendix}).  Simulation evidence appears in
Sections~\ref{sec: ATE simulation main}–\ref{sec: survival simulation main}. Section ~\ref{sec:reghal-atmle} extends to selecting data-dependent working models from a growing sieve, yielding a regularised adaptive-TMLE (regHAL-ATMLE).  Section~\ref{sec: conclusion}-\ref{sec: discussion}
conclude the paper and discuss future directions.

\medskip
% --------------------------------------------------------------------
\section{Preliminaries}\label{sec:preliminary}
% --------------------------------------------------------------------
\vspace{-0.5em}

\paragraph{Highly Adaptive Lasso}
HAL fits any càdlàg function on \([0,1]^d\) under a sectional-variation
norm constraint, attaining the $o_p(n^{-1/4})$ $L_2(P_0)$ rate
\cite{benkeser2016highly,vanderlaan2017generally,bibaut2019fast,vanderlaan2023higherordersplinehighly}.  
Concretely, it finds empirical risk minimizer among the 
\[
  Q_{\beta}(x)=\beta_0+\sum_{j=1}^{p}\beta_j\,\phi_j(x),
  \qquad\|\beta\|_1\le C,
\]
where each $\phi_j$ is a $0^{\text{th}}$-order indicator
$\mathbf1\{x_s>t\}$ (or a higher-order spline) chosen on a data-adaptive
grid \cite{tibshirani1996regression,zou2005regularization}.  
In practice, the cross validated lasso with a sequence of penalization parameter $\lambda$'s is being used to yield the finite-dimensional \textbf{working model}
\[
  M_n=\bigl\{Q_{\beta_{cv}}\bigr\},
  \qquad\dim(M_n)=p_{cv} \ll p\ll\infty ,
\]
where $\beta_{cv}$ are those $\beta$'s that take non zero values in the cross validated fit. This working model will be the playground for our targeting steps. In addition, we can consider the working model induced by the undersmoothed HAL fit (i.e. the lasso fit with penalization $\lambda$ that is smaller than the cross validated choice $\lambda_{cv}$).

\paragraph{Targeted Maximum Likelihood Estimation (TMLE)}
Given a pathwise-differentiable parameter $\Psi:\mathcal M\to\RR^k$ and an
initial nuisance parameters fit $(\hat Q_n,\hat g_n,\dots)$, TMLE finds
$Q_{n,\text{tmle}}$ along a least-favourable submodel so that
the empirical mean of the EIC vanishes
\cite{vanderlaan2011targeted,vanderlaan2018targeted}:
\[
  P_n D^{*}\!\bigl(Q_{n,\text{tmle}},\hat g_n,\dots\bigr)=0,
  \quad\Longrightarrow\quad
  \sqrt n\!\bigl[\Psi(\!Q_{n,\text{tmle}}\!) - \Psi(P_0)\bigr]
  \overset d\longrightarrow
  \NN\!\bigl(0,\Var_{P_0}[D^{*}(P_0)]\bigr).
\]

\paragraph{Working-model influence curves}
Inside $M_n$ the (parametric) EIC for $\Psi$ has the form
$
  D^{M_n,*}(P)=\gamma^\top S^\beta,
$
with $S^\beta$ the $p_{cv}$-vector of HAL scores and
$\gamma \in\RR^{p_{cv}}$ the coefficient that solves
$\gamma^\top I_P(\beta)=\partial_\beta\Psi(Q_\beta)$
\cite[Ch.~8]{vanderlaan2018targeted}, where $I_P(\beta)$ is the corresponding parametric fisher information matrix.

The difficulty is that the empirical Fisher matrix
$I_{P_n}(\beta)$ is often ill-conditioned (or even singular) because
many HAL basis functions are nearly collinear.  A classical fix is to
insert a small ridge penalization before inverting, replacing $I_{P_n}^{-1}$ by
$(I_{P_n}+\eta I)^{-1}$ \cite{carroll1998sandwich,tsiatis2006semiparametric};
in practice this has been used mainly for \emph{variance estimation} rather
than for the targeting step itself.  We revisit this idea
(Section~\ref{sec: EIC approximation}) for targeting step and compare it with a lasso‐projection
alternative that shrinks only those score directions relevant to the
chosen parameter \cite{vanderlaan2021efficient}.

\section{Problem Setup and Methodology}\label{sec: set up methodology}
We are interested in a smooth functional $\Psi$. Our target parameter of interest is $\Psi(Q_0)$, where the $Q_0$ is the relevant part of $P_0$ involved in $\Psi$. In addition, let's have additional nuisance part $g_0$ of $P_0\in M_{np}$.
We have $n$ iid samples from $P_0$ as $O_1,...O_n$. 

To construct a working model, we can first use the cross validated HAL to fit the $Q_0$ as $Q_n$. Then, we have a working model $M_n$ with $p$ parameters, induced by the active HAL basis. All of our HAL basis involved in $Q_n$ is within the $M_n$. Note that we can also naturally construct larger working models $M_n$  by including additional active basis in undersmoothed HAL fits.

We project the smooth functional as $\Psi_{M_n}(P)=\Psi(P_{M_n})$, where $P_{M_n}$ is the KL divergence projection of $P$ onto the $M_n$.
This defines our projected target parameter, $\Psi(P_{0,M_n})$.
We can restrict ourselves in the parametric working model $M_n$ and estimate $\Psi(P_{0,M_n})$ by updating the initial fit $Q_n$.  
Since we stay in the parametric model, this means we only need to update the coefficients. Note that our initial cross validated HAL estimate $Q_n$ may yield some zero coefficients in front of some basis involved in the $M_n$ when a larger working model is used.

Besides using the MLE in the $M_n$ as relaxed HAL to update the coefficients, we propose two ways to update the coefficients of $Q_n$ within $M_n$, based on two different approximations of the working model's parametric EIC (discussed in details in later sections).
Once the approximated parametric EIC within the model $M_n$ is obtained, we use the following general algorithm to construct targeted updates of initial HAL fits and yield our regularized HAL TMLE as regHAL-TMLEs :

\begin{algorithm}[H]
\caption{HAL Updating Algorithm Based on Approximated EIC (One Dimensional Target Parameter)}
\begin{algorithmic}[1]\label{1D regHAL-TMLE algorithm}
    \STATE \textbf{Cross-Validated HAL Fit:} Compute cross validated $Q_{n}$.
    \STATE \textbf{Working Model Definition:} Define the HAL working model $Q_{\beta}$.
    \STATE \textbf{Parameter Initialization:} Initialize $\beta$ based on $Q_{n}$, incorporating zero-padding if using an undersmoothed HAL model.
    \IF{using the projection-based method}
        \STATE \textbf{Additional Nuisance Fit:} Fit any extra nuisance parameters.
    \ENDIF
    \STATE \textbf{EIC Computation:} Derive the approximated Efficient Influence Curve (EIC)
    \[
    D^*_{n,\beta,Q}(O_i) = \alpha_n^T\,\text{scores}_{\beta}(O_i)
    \]
    at the current $\beta$, where $\text{scores}_{\beta}$ are the derivatives of log likelihood or negative loss w.r.t $\beta$.
    \STATE \textbf{Update Direction:} Determine the update direction for $\beta$ as
    \[
    \textit{Update} = \operatorname{sign}(P_nD^*_{n,\beta,Q})\,\alpha_n.
    \]
    \STATE \textbf{Parameter Update:} Update $\beta$ using gradient descent in the direction of $\textit{Update}$ with a small step size.
    \STATE \textbf{Stopping Criterion:} Repeat steps 7--9 until
    \[
    P_nD^*_{n,\beta,Q} < \frac{\operatorname{se}(D^*_{n,\beta,Q})}{\sqrt{n}\log(n)}.
    \]
\end{algorithmic}
\end{algorithm}

\paragraph{Remark: } For multi dimensional target parameter case, the algorithm is presented in appendix \ref{sec: reghal algorithm multi dimensional target}. 

\section{EIC Approximation Methods}\label{sec: EIC approximation}
We now discuss the two approximations of the working model's parametric EIC as $\alpha_n^T\,\text{scores}_{\beta}(O)$ used in Algorithm \ref{1D regHAL-TMLE algorithm}.

\subsection{Delta Method Based EIC Approximation}
Each $Q$ in the parametric model $M_n$ can be indexed by the coefficients $\beta$ as $Q(\beta)$.
Let the $S^{\beta}_{ij}$ be the score of $\beta_j$ of the log likelihood $l(O_i)$ for observation $O_i$ at $\beta$, and $S^{\beta}$ be the $n$ by $p$ score matrix at $\beta$ with entries $S^{\beta}_{ij}$. The $j$th column of $S^{\beta}$ is denoted as $S^{\beta}_{*,j}$ and the $i$th row as $S^{\beta}_{i,*}$.

The parametric EIC for $\Psi(Q(\beta))$ defined within this parametric model $M_n$ is:
\begin{align}
    D^*_{\beta}(O_i)=\{\frac{\partial{\Psi(Q(\beta))}}{\partial{\beta}}*I^{-1}_{P_{\beta}}(\beta)\}*(S^{\beta}_{i,*})^T
\end{align}
Here the $I_{P_{\beta}}(\beta)$ is the Fisher information matrix at $\beta$, where the $i,j$th entry is $E_{P_{\beta}}(S^{\beta}_i(O)S^{\beta}_j(O))$. Note, here $S^{\beta}_i(O)$ just means the score of log likelihood of $O$ w.r.t. $\beta_i$ at $\beta$.

Since $D^*_{\beta}(O_i)$ is linear in $S^{\beta}_{i,*}$,  in the TMLE procedure, $\{\frac{\partial{\Psi(Q(\beta))}}{\partial{\beta}}*I^{-1}_{P_{\beta}}(\beta)\}$ is our direction (for example, see Step 8 in Algorithm \ref{1D regHAL-TMLE algorithm}) to update the current $\beta$ to solve $D^*_{\beta}(O_i)$.
While $\frac{\partial{\Psi(Q(\beta))}}{\partial{\beta}}$ can be calculated or approximated easily,  $I^{-1}_{P_{\beta}}(\beta)$ is difficult to compute; therefore, usually $\{\frac{\partial{\Psi(Q(\beta))}}{\partial{\beta}}*I^{-1}_{P_n}(\beta)\}$ is used instead.
However, because HAL only aims to fit $Q_0$ well in the $L_2$ sense under the global sectional variation norm constraint, the selected parametric model $M_n$ may fail to achieve an optimal bias-variance trade-off using its parametric EIC for $\Psi$. In fact, with the built in collinearity of the HAL basis from construction, $I_{P_n}(\beta)$ may not be invertible.

% Thus,  $\{\frac{\partial{\Psi(Q(\beta))}}{\partial{\beta}}*I^{-1}_{P_n}(\beta)\}$ can be a bad estimator of $\{\frac{\partial{\Psi(Q(\beta))}}{\partial{\beta}}*I^{-1}_{P_{\beta}}(\beta)\}$

% \subsubsection{Regularization Blinded from Target Parameters}
To combat this issue -- although it may seem straightforward -- regularization blinded from the target parameters may not be sufficient. For example, 
one can consider PCA decomposition before the inverse, but this may completely ignore some variability important for $\frac{\partial{\Psi(Q(\beta))}}{\partial{\beta}}$. In this sense, ridge regularization (adding $\delta$ to the diagonal before taking inverse) offers a more suitable solution. Note that ridge regularization  equally reduces the variability in all directions~\cite{carroll1998sandwich,tsiatis2006semiparametric}.
For multidimensional target parameters, such regularization may not adapt to different directions (specified as different partial derivatives. While this may not cause significant issues for multidimensional target parameters with closely related elements (such as the survival curve), the performance can degrade when the elements differ more substantially (in how they depend on the nuisance parameters). For those more challenging scenarios, even when it works for one direction of the target parameter, it can perform suboptimally in the remaining directions.
% So, while the regularization may work for one direction and thus one target parameter, it may be very bad for another target parameter.

\subsection{Projection-Based EIC Approximation}
\label{sec:Proj-based}

Our second approximation method extends the general HAL-based EIC approximation framework \cite[Ch.~8]{vanderlaan2018targeted}  to the \emph{finite-dimensional working model} implied by the selected HAL basis. Rather than projecting onto a nonparametric space, we project an initial gradient (or a simpler influence function) onto the \emph{linear span} of the score functions associated with the HAL-selected basis coefficients under lasso type regularization. This yields a convenient, computationally stable approximation to the parametric EIC, even when the dimension of the selected basis is large or when collinearity issues arise. Compared with relaxed HAL or regularization based on the delta method, this HAL-based EIC approximation projects onto a slightly smaller tangent subspace, where the additional bounded variation assumption regularizes the potentially unstable components (such as $I^{-1}_{P_\beta}(\beta)$) and enables the HAL fit to bypass a matrix inverse.

Without regularizing the projection, for a gradient or an influence function $D$ (not necessarily the canonical gradient $D^*$), the parametric EIC $D^*_{\beta}(O_i)$ is represented by:
\begin{align}
    D^*_{\beta}(O_i)= \alpha_0^T*(S^{\beta}_{i,*})^T,
\end{align}
where $\alpha_0=\textit{argmin}_{\alpha}{P_{\beta}\bigg(D(O)-\sum_j\alpha_j S_{j}(O)\bigg)^2}$ and $S_{j}(O)$ is the score of $\beta_j$ of the log likelihood $l(O)$.
The finite sample approximation is given as 
\begin{align}
     D^*_{n,\beta}(O_i)=\alpha_n^T*(S^{\beta}_{i,*})^T,
\end{align}
where $\alpha_n=\textit{argmin}_{\alpha}{P_n\bigg(D(O)-\sum_j\alpha_j S_j(O)\bigg)^2}=I^{-1}_{P_n}(\beta) (S^{\beta})^T \Bar{D}$.
Here, $\Bar{D}=(D(O_1),D(O_2),...,D(O_n))^T$.
So,
\begin{align}
     D^*_{n,\beta}(O_i)=\{\Bar{D}^TS^{\beta}I^{-1}_{P_n}(\beta)\}*(S^{\beta}_{i,*})^T
\end{align}
Thus, we can estimate the direction of update $\{\frac{\partial{\Psi(Q(\beta))}}{\partial{\beta}}*I^{-1}_{P_{\beta}}(\beta)\}$  by $\{\Bar{D}^TS^{\beta}I^{-1}_{P_n}(\beta)\}$ when $I^{-1}_{P_n}(\beta)$ is stable with no severe collinearity among basis.

% \subsubsection{Regularization Adapted to the Target parameter}

But since we are essentially finding the closest $L_2(P_n)$ projection of $D(O)$ on the span of HAL scores as $ D^*_{n,\beta}(O)$, we can add some regularization to this projection, like a small $l1$ norm control, to get the regularized projection as $\sum_j\alpha_{n,j}^{reg}*S_j(O)$. And thus use the $\alpha_n^{reg,T}$ as the direction of update instead of the $\alpha_n^T$ which involves $I^{-1}_{P_n}(\beta)$. 

Such small regularization can be the key driver for favorable finite sample performance.
When interested in multiple target parameters at the same time, this lasso type regularization can really adapt to different directions (specified as different partial derivatives) as it fits a separate lasso projection for each target parameter of interest and thus regularize different part of the current fit.
In addition, the regularization level can be fully data-dependent (through a cross-validated procedure such as \texttt{cv.glmnet} with lasso or elastic net).

\subsubsection{Projection Regularization's relationship to collaborative TMLE}

There is a very natural connection to the collaborative TMLE \cite{ju2019scalable} of our projection based targeting in the sense that the clever co-variates existed in the TMLE is naturally regulated by the features (the HAL basis in the selected working model) that are meaningful for the relevant part of $P$ as $Q(P)$ related to the target parameter of interest $\Psi(P)=\Psi(Q(P))$.

In specific, one can estimate the nuisance part in the initial gradient or canonical gradient $D$ ignorant of the relevant part $Q(P)$ but then when projecting $D$ onto the score space related to the HAL working model, the nuisance part is naturally, through clever co-variates, regulated by the features as HAL basis that are important for the estimation of $Q(P)$.

For example, in the classical ATE set up of data $O=(W,A,Y)$, our scores for $Q(P)$ HAL working model takes the form of $\phi_j(Y-Q(W,A))$, where $Q(W,A)=E(Y|A,W)$, and our canonical gradient part relevant to $Q(P)$ is $(\frac{A}{g(A|W)}-\frac{1-A}{g(A|W)})(Y-Q(W,A))$, where $g(A|W)=P(A|W)$. Thus, one can first estimate the $g$ as $\hat{g}$ ignorant of the $Q(W,A)$ and estimate the $Q$ using HAL as $\hat{Q}$ ignorant of $g$, and one get $\hat{D}=(\frac{A}{\hat{g}(A|W)}-\frac{1-A}{\hat{g}(A|W)})(Y-\hat{Q}(W,A))$

So, when doing the projection, one is just trying to span the $\frac{A}{\hat{g}(A|W)}-\frac{1-A}{\hat{g}(A|W)}$ by the $\phi_j(A,W)$ involved in $\hat{Q}(W,A)$. Thus, one can think that linear span as the estimate of the clever co-variate itself and thus in our projection based targeting, the clever co-variate is collaboratively regularized before it is used for the targeting purpose.

\section{Theoretical Guarantees}\label{sec:theory}
In this section, we closely followed the theorems and the proofs in \cite{vdl_adml_2023}. For clarity and completeness, we adapt them into our notations and set ups below.

\subsection{Equality of $D_{M_n}^{*}(P)$ and $D^{M_n,*}(P)$}
\label{subsec:EIC_equality}

We use the superscript $\Psi^{M_n}: M_n\to\RR$  to represent the original parameter applied within $P \in M_n$, and the subscript $\Psi_{M_n}=\Psi\circ\Pi_{M_n}: M_{np}\to\RR$ to represent the projection parameter, which is defined for all $P \in M_{np}$ but resulting in the same value range. 
In fact, $\Psi^{M_n}(P)= \Psi_{M_n}(P) = \Psi(P)$ for all $P\in M_n$. 
% Let restriction of $\Psi$ to the working model as
      % \(\Psi^{M_n}: M_n\to\RR,\ \textit{with }  \Psi^{M_n}(P)=\Psi(P)\) $\forall P\in M_n$. 

\begin{assumption}\label{as:proj-diff}
For the distribution $P$ under consideration we assume:
\begin{enumerate}[label=(\alph*)]
\item \textbf{Projection differentiability}\,: \\ 
      $\Pi_{M_n}$ is Gâteaux differentiable at $P$ and  
      \(\displaystyle
        \Bigl\{\,\dot Q =
          \tfrac{d}{d\varepsilon}\Pi_{M_n}(P_\varepsilon)\!\mid_{0}
          : S_{P_\varepsilon}\in T_P(\mathcal M)\Bigr\}
        = T_P(M_n)\).
\item \textbf{Smoothness of $\Psi$ on $M_n$}\,:  \\
      $\Psi$ is pathwise differentiable on $M_n$; denote its canonical
      gradient by \(D^{M_n,*}(Q)\) for each $Q\in M_n$.
\end{enumerate}
\end{assumption}

\begin{theorem}[Coincidence of influence curves on $M_n$]
\label{thm:EIC_equality_proj_restrict}
Let Assumption~\ref{as:proj-diff} hold and suppose that the reference
distribution lies \emph{in} the working model:
\(P\in M_n\).
Then
\[
      D_{M_n}^{*}(P)
      \;=\;
      D^{M_n,*}(P)
      \qquad\text{in }L^{2}(P),
\]
where
\begin{itemize}
\item $D_{M_n}^{*}(P)$ is the efficient influence curve of the
      projected parameter \( \Psi_{M_n}\) in the full model
      $\mathcal M$;
\item $D^{M_n,*}(P)$ is the efficient influence curve of the
      restricted map \(\Psi^{M_n}\) computed inside $M_n$.
\end{itemize}
\end{theorem}

\begin{remark}
If \(P\notin M_n\) the two objects generally differ:  
$\Psi_{M_n}$ is still well defined for every $P$, but
$\Psi^{M_n}$ (and thus \(D^{M_n,*}\)) is only defined on~$M_n$.
\end{remark}

\subsection{TMLE for working model induced target parameter}
Define $R_{M_n}(P,P_0)=\Psi_{M_n}(P)-\Psi_{M_n}(P_0)+P_0D^*_{M_n}(P)$ be the exact remainder, where $D^*_{M_n}(P)$ is the EIC of $\Psi_{M_n}$ at $P$ relative to $M_{np}$.

\begin{theorem}\label{thm:TMLE working model}
    Let $P_{M_n}^*$ denote our regHAL-TMLE that solves $P_nD^{M_n,*}(P_{M_n}^*)=P_nD^*_{M_n}(P_{M_n}^*)=o_p(n^{-1/2})$.
    Assume
    \begin{enumerate}[label=\textnormal{(A.\arabic*)},
                  ref=A.\arabic*,
                  leftmargin=2.5em,   % ← aligns nicely under “Assume”
                  itemsep=0pt]        % ← tight vertical spacing
        \item $D^*_{M_n}(P_{M_n}^*),D^*_{M_n}(P_0)$ lie in $P_0$-Donsker class with high probability
        \item $P_0\big(D^*_{M_n}(P_{M_n}^*)-D^*_{M_n}(P_0)\big)^2=o_p(1)$
    \end{enumerate}
    Then 
    \begin{align*}
    &\Psi_{M_n}(P_{M_n}^*)-\Psi_{M_n}(P_0)=P_nD^*_{M_n}(P_0)+o_p(n^{-1/2})
\end{align*}
\end{theorem}

\subsection{TMLE for original target parameter}
Define $R(P,P_0)=\Psi(P)-\Psi(P_0)+P_0D^*(P)$ be the exact remainder, where $D^*(P)$ is the EIC of $\Psi$ at $P$ relative to $M_{np}$.
\begin{theorem}\label{thm:TMLE}
If, in addition, there exists $M_0$ that contains $P_0$ such that 
    \begin{enumerate}[label=\textnormal{(B.\arabic*)},
                      ref=B.\arabic*,
                      leftmargin=2.5em,   % ← aligns nicely under “Assume”
                      itemsep=0pt]        % ← tight vertical spacing
        \item $D^*_{M_0}(P_0)$ lies in $P_0$-Donsker class with high probability
        \item $P_0\big(D^*_{M_n}(P_0)-D^*_{M_0}(P_0)\big)^2=o_p(1)$
    \end{enumerate}
and
    \begin{enumerate}[label=\textnormal{(C.\arabic*)},
                      ref=C.\arabic*,
                      leftmargin=2.5em,   % ← aligns nicely under “Assume”
                      itemsep=0pt]        % ← tight vertical spacing
        \item $\Psi(P_{0,M_n})-\Psi_{M_0}(P_{0,M_n})=o_p(n^{-1/2})$
        \item $(P_{0,M_n}-P_0)(D^*_{M_0}(P_{0,M_n})-D^*_{M_n}(P_{0,M_n}))=o_p(n^{-1/2})$
        \item $P_0D^*_{M_n}(P_{0,M_n})=o_p(n^{-1/2})$
    \end{enumerate}
    Then,
    \begin{align*}
    &\Psi_{M_n}(P_{M_n}^*)-\Psi(P_0)=P_nD^*_{M_0}(P_0)+o_p(n^{-1/2})
    \end{align*} 
\end{theorem}

\subsection{Verification of Regularity Conditions (A.1)–(C.3) under HAL}\label{sec:verify_all}

We now check that, under the usual HAL‐sieve approximability and convergence
rates, all of the technical assumptions in Theorems~\ref{thm:TMLE working model}
and~\ref{thm:TMLE}—namely (A.1)–(A.2), (B.1)–(B.2) and (C.1)–(C.3)—are satisfied.

\begin{assumption}[HAL‐sieve approximation]\label{as:hal_sieve_full}
Let \(Q_0\) and every nuisance (e.g.\ propensity score \(g_0\), censoring
hazards) lie in the class of càdlàg functions on \([0,1]^d\) with sectional‐
variation norm bounded by \(M\).  Let \(M_n\) be the finite‐dimensional
HAL sieve selected by (possibly under‐smoothed) cross‐validation, and
\(\hat Q_n\in M_n\) its order-\(k\) HAL‐MLE.  Then for some
\(r>1/4\),
\[
  \|\hat Q_n - Q_0\|_\infty = o_P(1),
  \quad
  \|\hat Q_n - Q_0\|_{L^2(P_0)} = O_P(n^{-r}),
\]
and the same sup‐norm and \(L^2\) rates hold for every other HAL‐based
nuisance estimator \cite{vanderlaan2023higherordersplinehighly}.  Denote the KL‐projection of
\(Q_0\) onto \(M_n\) by \(Q_{0,M_n}=\Pi_{M_n}(Q_0)\).
\end{assumption}

\begin{theorem}\label{prop:verify_AtoC}
Under Assumption~\ref{as:hal_sieve_full}, (A.1)-(A.2),(B.1)-(B.2),(C.1)-(C.3) hold.
\end{theorem}

% Section 6: ATE Results
\section{regHAL-TMLE for Average Treatment Effects}\label{sec: ATE simulation main}

We evaluate our regularized targeting approaches through comprehensive simulations for average treatment effect (ATE) estimation, comparing projection-based and delta-method targeting against relaxed HAL baselines. Implementation and statistical details appear in Appendix~\ref{sec:regHAL-ate-details}.

\subsection{Simulation Design}
We implement both proposed methods using weak regularization: projection-based targeting with $\lambda=10^{-5}$ for Lasso projection, and delta-method targeting with ridge parameter $\eta=10^{-6}$. Gradient step size is $10^{-4}$ for both approaches. We evaluate performance within cross-validated HAL working models and undersmoothed variants (using the 10th largest $\lambda < \lambda_{cv}$ or smallest available).

We use two data generating processes to test method robustness: DGP~1 represents standard conditions without positivity violations, while DGP~2 introduces practical positivity violations through extreme propensity scores (details in Appendix~\ref{sec: DGP1}--\ref{sec: DGP2}).

\subsection{Results Under Standard Conditions}
Table~1,2 presents results under DGP~1 using 500 replications across sample sizes $n \in \{500, 1000, 1500, 2000\}$. Below are some key findings.

\textbf{Superior bias reduction} Projection-based targeting achieves 50--85\% bias reduction versus relaxed HAL across all scenarios. At $n=500$, absolute bias drops from 0.0421 (relaxed) to 0.0058 (projection).

\textbf{MSE improvements} Projection targeting reduces MSE by 30--40\% compared to relaxed HAL and 15--25\% versus delta-method targeting. Performance advantages persist under undersmoothing, indicating effective regularization of expanded basis sets.

\textbf{Coverage properties} Projection-based confidence intervals achieve near-nominal coverage using nonparametric EIC construction, reaching 95\% at $n=1000$ and maintaining this level with larger samples. Delta-method targeting shows systematic under-coverage for delta-specific intervals.

\begin{table}[H]
\centering
\centering
\resizebox{\ifdim\width>\linewidth\linewidth\else\width\fi}{!}{
\fontsize{8}{10}\selectfont
\begin{tabular}[t]{cccrrrrrrrr}
\toprule
Estimator & n & Targeting & Abs\_Bias & Std\_Err & MSE & Cov NP (\%) & Cov Proj (\%) & Cov Proj CV (\%) & Cov Delta (\%) & Oracle Cov (\%)\\
\midrule
 & 500 & Relaxed & 0.0421 & 0.1595 & 0.027072 & 80.00 (0.439) & 65.00 (0.343) & 61.50 (0.309) & 63.50 (0.320) & 97.00 (0.625)\\

 & 500 & Projection & 0.0058 & 0.1309 & 0.017086 & 89.50 (0.445) & 85.50 (0.368) & 80.50 (0.343) & 76.50 (0.306) & 94.50 (0.513)\\

 & 500 & Delta-method & 0.0529 & 0.1388 & 0.021982 & 87.50 (0.445) & 77.00 (0.367) & 74.00 (0.341) & 66.50 (0.307) & 93.50 (0.544)\\
\cmidrule(lr){2-11}

 & 1000 & Relaxed & 0.0067 & 0.0860 & 0.007400 & 91.50 (0.312) & 87.00 (0.249) & 85.00 (0.233) & 86.00 (0.229) & 93.00 (0.337)\\

 & 1000 & Projection & 0.0067 & 0.0775 & 0.006024 & 95.00 (0.314) & 92.50 (0.265) & 91.50 (0.256) & 85.50 (0.220) & 94.50 (0.304)\\

 & 1000 & Delta-method & 0.0304 & 0.0834 & 0.007852 & 91.50 (0.314) & 88.00 (0.265) & 87.50 (0.255) & 80.00 (0.220) & 93.00 (0.327)\\
\cmidrule(lr){2-11}

 & 1500 & Relaxed & 0.0021 & 0.0668 & 0.004448 & 96.50 (0.256) & 87.00 (0.206) & 81.50 (0.198) & 80.50 (0.186) & 97.50 (0.262)\\

 & 1500 & Projection & 0.0025 & 0.0659 & 0.004333 & 97.50 (0.258) & 89.50 (0.218) & 89.00 (0.214) & 78.00 (0.179) & 97.50 (0.258)\\

 & 1500 & Delta-method & 0.0183 & 0.0683 & 0.004972 & 94.50 (0.258) & 86.00 (0.218) & 85.50 (0.213) & 76.50 (0.179) & 95.00 (0.268)\\
\cmidrule(lr){2-11}

 & 2000 & Relaxed & 0.0037 & 0.0577 & 0.003329 & 97.50 (0.222) & 86.50 (0.177) & 85.50 (0.171) & 83.50 (0.162) & 98.00 (0.226)\\

 & 2000 & Projection & 0.0085 & 0.0570 & 0.003303 & 96.50 (0.222) & 89.50 (0.185) & 88.50 (0.180) & 82.50 (0.158) & 96.50 (0.223)\\

\multirow{-12}{*}{\centering\arraybackslash HAL-TMLE} & 2000 & Delta-method & 0.0196 & 0.0578 & 0.003708 & 94.00 (0.222) & 87.00 (0.185) & 86.00 (0.179) & 80.50 (0.158) & 94.50 (0.227)\\
\bottomrule
\end{tabular}}
\caption{Simulation results for HAL-TMLE. DGP 1. CV selected}
\end{table}

\begin{table}[H]
\centering
\centering
\resizebox{\ifdim\width>\linewidth\linewidth\else\width\fi}{!}{
\fontsize{8}{10}\selectfont
\begin{tabular}[t]{cccrrrrrrrr}
\toprule
Estimator & n & Targeting & Abs\_Bias & Std\_Err & MSE & Cov NP (\%) & Cov Proj (\%) & Cov Proj CV (\%) & Cov Delta (\%) & Oracle Cov (\%)\\
\midrule
 & 500 & Relaxed & 0.0297 & 0.1626 & 0.027149 & 79.52 (0.436) & 66.27 (0.343) & 61.45 (0.301) & 66.27 (0.330) & 96.39 (0.637)\\

 & 500 & Projection & 0.0102 & 0.1363 & 0.018566 & 88.55 (0.445) & 85.54 (0.379) & 80.72 (0.349) & 74.10 (0.307) & 94.58 (0.534)\\

 & 500 & Delta-method & 0.0490 & 0.1474 & 0.024001 & 84.94 (0.445) & 77.11 (0.378) & 71.08 (0.350) & 65.66 (0.309) & 94.58 (0.578)\\
\cmidrule(lr){2-11}

 & 1000 & Relaxed & 0.0150 & 0.0851 & 0.007359 & 92.86 (0.312) & 90.00 (0.250) & 90.00 (0.231) & 90.00 (0.229) & 94.29 (0.334)\\

 & 1000 & Projection & 0.0165 & 0.0814 & 0.006805 & 94.29 (0.315) & 92.86 (0.270) & 92.86 (0.262) & 84.29 (0.218) & 94.29 (0.319)\\

 & 1000 & Delta-method & 0.0419 & 0.0865 & 0.009125 & 92.86 (0.315) & 87.14 (0.270) & 82.86 (0.262) & 74.29 (0.218) & 92.86 (0.339)\\
\cmidrule(lr){2-11}

 & 1500 & Relaxed & 0.0035 & 0.0671 & 0.004490 & 95.00 (0.254) & 83.50 (0.203) & 81.50 (0.187) & 83.00 (0.191) & 97.00 (0.263)\\

 & 1500 & Projection & 0.0025 & 0.0660 & 0.004335 & 97.50 (0.258) & 89.50 (0.220) & 88.00 (0.209) & 78.00 (0.182) & 97.00 (0.259)\\

 & 1500 & Delta-method & 0.0191 & 0.0694 & 0.005163 & 94.00 (0.258) & 85.50 (0.220) & 81.50 (0.209) & 76.00 (0.181) & 95.00 (0.272)\\
\cmidrule(lr){2-11}

 & 2000 & Relaxed & 0.0007 & 0.0573 & 0.003268 & 96.50 (0.219) & 87.50 (0.174) & 84.50 (0.163) & 86.00 (0.166) & 96.50 (0.225)\\

 & 2000 & Projection & 0.0085 & 0.0570 & 0.003303 & 96.50 (0.222) & 90.00 (0.186) & 88.00 (0.177) & 84.00 (0.160) & 96.50 (0.223)\\

\multirow{-12}{*}{\centering\arraybackslash HAL-TMLE} & 2000 & Delta-method & 0.0195 & 0.0582 & 0.003749 & 94.50 (0.222) & 87.00 (0.186) & 85.00 (0.176) & 80.00 (0.160) & 95.00 (0.228)\\
\bottomrule
\end{tabular}}
\caption{Simulation results for HAL-TMLE. DGP 1. Undersmoothed.}
\end{table}

\subsection{Performance Under Positivity Violations}
Table~3,4 examines performance under DGP~2's challenging positivity violation scenario.

\textbf{Dramatic improvements} Projection-based targeting shows 80\% bias reduction ($n=500$: 0.2056 $\to$ 0.0416) while maintaining reasonable coverage (92--95\%) even under severe positivity violations. Relaxed HAL and delta-method targeting suffer substantial coverage degradation, with delta-specific intervals achieving only 28--52\% coverage in challenging conditions.

The projection method's advantages increase with sample size, suggesting effective bias-variance trade-off management through parameter-specific regularization rather than uniform shrinkage across all score directions.

\begin{table}[H]
\centering
\centering
\resizebox{\ifdim\width>\linewidth\linewidth\else\width\fi}{!}{
\fontsize{8}{10}\selectfont
\begin{tabular}[t]{cccrrrrrrrr}
\toprule
Estimator & n & Targeting & Abs\_Bias & Std\_Err & MSE & Cov NP (\%) & Cov Proj (\%) & Cov Proj CV (\%) & Cov Delta (\%) & Oracle Cov (\%)\\
\midrule
 & 500 & Relaxed & 0.2056 & 0.1659 & 0.069682 & 85.00 (0.999) & 68.00 (0.642) & 57.00 (0.500) & 52.00 (0.400) & 74.00 (0.650)\\

 & 500 & Projection & 0.0416 & 0.2402 & 0.059117 & 92.50 (0.958) & 73.00 (0.629) & 63.00 (0.537) & 53.50 (0.381) & 95.00 (0.941)\\

 & 500 & Delta-method & 0.1947 & 0.2151 & 0.083918 & 88.00 (1.062) & 74.00 (0.697) & 61.00 (0.541) & 52.00 (0.392) & 90.00 (0.843)\\
\cmidrule(lr){2-11}

 & 1000 & Relaxed & 0.1978 & 0.1244 & 0.054507 & 83.00 (0.715) & 58.00 (0.462) & 53.50 (0.423) & 32.50 (0.289) & 62.00 (0.488)\\

 & 1000 & Projection & 0.0307 & 0.1795 & 0.032999 & 95.50 (0.713) & 77.00 (0.472) & 74.00 (0.435) & 49.00 (0.272) & 96.00 (0.704)\\

 & 1000 & Delta-method & 0.1911 & 0.1330 & 0.054117 & 88.00 (0.746) & 62.00 (0.489) & 55.50 (0.436) & 28.00 (0.289) & 70.00 (0.521)\\
\cmidrule(lr){2-11}

 & 1500 & Relaxed & 0.2027 & 0.1086 & 0.052804 & 73.50 (0.585) & 43.00 (0.378) & 40.00 (0.357) & 21.00 (0.231) & 55.00 (0.426)\\

 & 1500 & Projection & 0.0302 & 0.1479 & 0.022687 & 93.00 (0.593) & 77.50 (0.398) & 76.50 (0.380) & 56.00 (0.218) & 94.00 (0.580)\\

 & 1500 & Delta-method & 0.2031 & 0.1072 & 0.052679 & 79.00 (0.608) & 48.50 (0.399) & 43.50 (0.373) & 21.50 (0.227) & 56.50 (0.420)\\
\cmidrule(lr){2-11}

 & 2000 & Relaxed & 0.0686 & 0.1183 & 0.018643 & 90.00 (0.466) & 77.00 (0.342) & 75.50 (0.326) & 64.50 (0.247) & 90.00 (0.464)\\

 & 2000 & Projection & 0.0265 & 0.1178 & 0.014519 & 95.00 (0.479) & 82.00 (0.362) & 80.00 (0.346) & 65.00 (0.230) & 94.50 (0.462)\\

\multirow{-12}{*}{\centering\arraybackslash HAL-TMLE} & 2000 & Delta-method & 0.0707 & 0.2146 & 0.050840 & 86.00 (0.491) & 70.50 (0.366) & 68.50 (0.351) & 54.00 (0.240) & 98.50 (0.841)\\
\bottomrule
\end{tabular}}
\caption{Simulation results for HAL-TMLE under DGP 2 with CV-selected working model.}
\end{table}

\begin{table}[H]
\centering
\centering
\resizebox{\ifdim\width>\linewidth\linewidth\else\width\fi}{!}{
\fontsize{8}{10}\selectfont
\begin{tabular}[t]{cccrrrrrrrr}
\toprule
Estimator & n & Targeting & Abs\_Bias & Std\_Err & MSE & Cov NP (\%) & Cov Proj (\%) & Cov Proj CV (\%) & Cov Delta (\%) & Oracle Cov (\%)\\
\midrule
 & 500 & Relaxed & 0.1985 & 0.1738 & 0.069422 & 87.72 (0.982) & 72.51 (0.644) & 57.31 (0.484) & 53.22 (0.427) & 76.02 (0.681)\\

 & 500 & Projection & 0.0339 & 0.2350 & 0.056052 & 92.40 (0.967) & 77.78 (0.655) & 65.50 (0.531) & 54.39 (0.396) & 94.74 (0.921)\\

 & 500 & Delta-method & 0.2115 & 0.2651 & 0.114612 & 89.47 (1.071) & 78.36 (0.725) & 63.16 (0.522) & 52.05 (0.419) & 95.32 (1.039)\\
\cmidrule(lr){2-11}

 & 1000 & Relaxed & 0.2118 & 0.1181 & 0.058706 & 79.82 (0.728) & 56.88 (0.470) & 50.46 (0.428) & 28.44 (0.294) & 53.21 (0.463)\\

 & 1000 & Projection & 0.0445 & 0.1824 & 0.034936 & 95.41 (0.730) & 76.15 (0.487) & 73.39 (0.447) & 49.54 (0.273) & 95.41 (0.715)\\

 & 1000 & Delta-method & 0.2040 & 0.1344 & 0.059521 & 84.40 (0.756) & 59.63 (0.498) & 52.29 (0.450) & 26.61 (0.285) & 66.06 (0.527)\\
\cmidrule(lr){2-11}

 & 1500 & Relaxed & 0.0691 & 0.1157 & 0.018090 & 91.00 (0.535) & 83.50 (0.395) & 78.00 (0.367) & 68.00 (0.275) & 91.00 (0.453)\\

 & 1500 & Projection & 0.0348 & 0.1441 & 0.021861 & 93.00 (0.582) & 83.00 (0.451) & 79.50 (0.418) & 62.00 (0.252) & 94.00 (0.565)\\

 & 1500 & Delta-method & 0.0662 & 0.2599 & 0.071612 & 88.50 (0.610) & 76.00 (0.466) & 72.50 (0.430) & 53.50 (0.266) & 96.50 (1.019)\\
\cmidrule(lr){2-11}

 & 2000 & Relaxed & 0.0321 & 0.1166 & 0.014558 & 91.50 (0.433) & 82.50 (0.352) & 79.50 (0.332) & 76.00 (0.304) & 94.00 (0.457)\\

 & 2000 & Projection & 0.0247 & 0.1192 & 0.014738 & 94.00 (0.474) & 85.50 (0.400) & 83.50 (0.383) & 70.00 (0.267) & 95.00 (0.467)\\

\multirow{-12}{*}{\centering\arraybackslash HAL-TMLE} & 2000 & Delta-method & 0.0310 & 0.2959 & 0.088065 & 87.00 (0.493) & 79.50 (0.411) & 75.50 (0.390) & 62.00 (0.283) & 96.00 (1.160)\\
\bottomrule
\end{tabular}}
\caption{Simulation results for HAL-TMLE under DGP 2 with undersmoothed working model.}
\end{table}

\subsection{Methodological Insights}
Projection-based targeting's consistent superiority stems from adaptive regularization that selectively shrinks HAL basis components contributing minimally to bias reduction for the specific target parameter. This targeted approach proves especially valuable under collinearity---common in HAL implementations---and challenging inference conditions where robust performance is most critical.

Comparison with standard TMLE (Appendix~\ref{appendix: standard TMLE}) shows comparable performance under normal conditions but substantial advantages under positivity violations, achieved without ad hoc truncation decisions.

% Section 7: Survival Results  
\section{regHAL-TMLE for Survival Curves}\label{sec: survival simulation main}

We demonstrate regHAL-TMLE scalability through survival curve estimation from right-censored data---a high-dimensional target parameter showcasing our methodology's advantages in computationally demanding scenarios. Implementation and statstical details appear in Appendix~\ref{sec:regHAL-survival-details}.

\subsection{High-Dimensional Targeting Challenges}
Survival curve/surface estimation in general can pose unique computational challenges that traditional HAL-TMLE approaches struggle to address:

\textbf{Exponential complexity} Standard methods require recursive clever covariate calculations at each time point and iteration, with computational burden scaling exponentially with grid size.

\textbf{Numerical instability} Numerical integration techniques involved can create mounting instability as iterations progress.

\textbf{Working model advantages} Our approach sidesteps these issues by operating entirely within HAL-implied working models, where each update modifies only coefficient vectors while preserving survival function monotonicity.

\subsection{Implementation}
We employ intensity-based HAL models using Poisson regression with repeated observations. Failure time hazard $\lambda^T(t)$ uses 100 percentiles of observed event times as knot points; censoring hazard $\lambda^C(t)$ uses 50 percentiles for efficiency. Undersmoothed models multiply the $L_1$ constraint by 1.61 to systematically expand working model complexity.

For survival curves evaluated at time grid $G$, we employ universal least favorable path updates (Algorithm~2, Appendix~\ref{sec: reghal algorithm multi dimensional target}) with step size $10^{-3}$ for both targeting approaches.

\subsection{Simulation Results}
Tables~5--6 present results using 500 replications across $n \in \{500, 1000, 1500, 2000\}$ under the data generating process described in Appendix~\ref{sec: survival DGP}.

\textbf{Relaxed HAL failure} Standard relaxed targeting exhibits severe bias persistence (absolute bias 0.30--0.44) and MSE above 0.13 across all sample sizes, with confidence interval coverage of only 3--23\%.

\textbf{Regularized success} Both projection-based and delta-method targeting achieve $\approx$98\% bias reduction (absolute bias 0.007--0.014) with MSE decreasing to 0.0001--0.0004. Coverage using regularized EIC constructions reaches 87--93\%, demonstrating effective distributional approximation.

\textbf{Method comparison} Projection-based targeting achieves $\approx$10\% lower MSE than delta-method targeting while maintaining comparable coverage properties. Both methods show remarkable consistency across CV-selected and undersmoothed working models.

\begin{table}[H]
\centering
\resizebox{\ifdim\width>\linewidth\linewidth\else\width\fi}{!}{%
\fontsize{8}{10}\selectfont
\begin{tabular}{lcccrrrrrrr}
\toprule
Estimator & n & Targeting     & Abs\_Bias & Std\_Err & MSE       & Cov NP (\%) & Cov Proj (\%) & Cov Proj CV (\%) & Cov Delta (\%) & Oracle Cov (\%)\\
\midrule
\multirow{12}{*}{Intensity-HAL-TMLE} 
  & 500  & Relaxed      & 0.3585  & 0.0551  & 0.189461  & --         & 23.81 (0.150) & 23.72 (0.148)  & 6.26 (0.113)  & 16.04 (0.216) \\
  & 500  & Projection   & 0.0137  & 0.0166  & 0.000367  & --         & 86.94 (0.063) & 67.77 (0.050)  & 90.28 (0.071) & 93.18 (0.065) \\
  & 500  & Delta-method & 0.0139  & 0.0169  & 0.000377  & --         & 87.63 (0.065) & 68.74 (0.050)  & 91.07 (0.070) & 93.26 (0.066) \\
\cmidrule(lr){2-11}
  & 1000 & Relaxed      & 0.4022  & 0.0470  & 0.239075  & --         & 20.19 (0.111) & 20.16 (0.110)  & 4.13 (0.056)  & 12.78 (0.184) \\
  & 1000 & Projection   & 0.0098  & 0.0117  & 0.000182  & --         & 88.53 (0.045) & 72.72 (0.038)  & 91.96 (0.050) & 92.77 (0.046) \\
  & 1000 & Delta-method & 0.0099  & 0.0118  & 0.000189  & --         & 88.90 (0.046) & 72.98 (0.038)  & 92.87 (0.050) & 93.16 (0.046) \\
\cmidrule(lr){2-11}
  & 1500 & Relaxed      & 0.4211  & 0.0564  & 0.263935  & --         & 18.31 (0.092) & 18.29 (0.091)  & 3.89 (0.040)  & 9.92 (0.221) \\
  & 1500 & Projection   & 0.0084  & 0.0099  & 0.000133  & --         & 87.51 (0.037) & 73.66 (0.032)  & 91.78 (0.041) & 91.89 (0.039) \\
  & 1500 & Delta-method & 0.0086  & 0.0103  & 0.000144  & --         & 87.49 (0.038) & 73.78 (0.032)  & 92.17 (0.041) & 92.32 (0.040) \\
\cmidrule(lr){2-11}
  & 2000 & Relaxed      & 0.4382  & 0.0331  & 0.285512  & --         & 17.08 (0.080) & 17.08 (0.080)  & 4.27 (0.029)  & 11.08 (0.130) \\
  & 2000 & Projection   & 0.0071  & 0.0084  & 0.000094  & --         & 88.38 (0.033) & 76.12 (0.028)  & 92.39 (0.036) & 91.12 (0.033) \\
  & 2000 & Delta-method & 0.0073  & 0.0087  & 0.000100  & --         & 88.25 (0.033) & 75.96 (0.028)  & 93.09 (0.036) & 91.98 (0.034) \\
\bottomrule
\end{tabular}%
}
\caption{Simulation results for Intensity HAL TMLE with CV-selected working model.}
\label{tab:sim_results_intensity_1d_cv}
\end{table}

\begin{table}[H]
\centering
\resizebox{\ifdim\width>\linewidth\linewidth\else\width\fi}{!}{%
\fontsize{8}{10}\selectfont
\begin{tabular}{lcccrrrrrrr}
\toprule
Estimator & n & Targeting     & Abs\_Bias & Std\_Err & MSE       & Cov NP (\%) & Cov Proj (\%) & Cov Proj CV (\%) & Cov Delta (\%) & Oracle Cov (\%)\\
\midrule
\multirow{12}{*}{Intensity-HAL-TMLE} 
  & 500  & Relaxed      & 0.3045  & 0.0598  & 0.137199  & --         & 24.76 (0.147) & 24.52 (0.144)  & 9.65 (0.142)   & 14.38 (0.235) \\
  & 500  & Projection   & 0.0144  & 0.0168  & 0.000400  & --         & 86.22 (0.065) & 66.45 (0.050)  & 90.71 (0.075)  & 92.17 (0.066) \\
  & 500  & Delta-method & 0.0149  & 0.0175  & 0.000435  & --         & 87.00 (0.068) & 67.44 (0.050)  & 91.71 (0.074)  & 92.25 (0.069) \\
\cmidrule(lr){2-11}
  & 1000 & Relaxed      & 0.3741  & 0.0370  & 0.204782  & --         & 20.41 (0.111) & 20.37 (0.110)  & 4.53 (0.066)   & 9.28 (0.145)  \\
  & 1000 & Projection   & 0.0104  & 0.0120  & 0.000200  & --         & 87.48 (0.046) & 71.09 (0.038)  & 91.57 (0.052)  & 92.07 (0.047) \\
  & 1000 & Delta-method & 0.0108  & 0.0124  & 0.000224  & --         & 87.83 (0.047) & 71.53 (0.038)  & 92.56 (0.053)  & 92.45 (0.049) \\
\cmidrule(lr){2-11}
  & 1500 & Relaxed      & 0.4041  & 0.0361  & 0.241252  & --         & 18.47 (0.092) & 18.47 (0.091)  & 4.25 (0.044)   & 8.04 (0.142)  \\
  & 1500 & Projection   & 0.0087  & 0.0101  & 0.000143  & --         & 86.65 (0.038) & 72.04 (0.032)  & 91.18 (0.043)  & 90.97 (0.039) \\
  & 1500 & Delta-method & 0.0092  & 0.0108  & 0.000167  & --         & 86.27 (0.038) & 72.14 (0.032)  & 91.89 (0.043)  & 91.47 (0.042) \\
\cmidrule(lr){2-11}
  & 2000 & Relaxed      & 0.4234  & 0.0298  & 0.265035  & --         & 16.84 (0.080) & 16.84 (0.080)  & 3.80 (0.030)   & 7.70 (0.117)  \\
  & 2000 & Projection   & 0.0074  & 0.0086  & 0.000101  & --         & 87.64 (0.033) & 74.37 (0.028)  & 91.75 (0.037)  & 90.83 (0.034) \\
  & 2000 & Delta-method & 0.0078  & 0.0091  & 0.000118  & --         & 87.26 (0.033) & 74.21 (0.029)  & 93.07 (0.037)  & 91.67 (0.036) \\
\bottomrule
\end{tabular}%
}
\caption{Simulation results for Intensity HAL TMLE with undersmoothed working model.}
\label{tab:sim_results_intensity_1d_undersmoothed}
\end{table}

\subsection{Computational and Methodological Implications}
Results demonstrate that working model selection alone is insufficient for complex target parameters---effective targeting requires careful EIC approximation respecting both parameter structure and working model geometry. 

The complete failure of relaxed HAL despite identical initial fits, combined with regularized methods' success, establishes regHAL-TMLE as essential for high-dimensional inference where traditional approaches face fundamental computational and statistical barriers. Our $O(p)$ per-iteration complexity scales naturally to larger time grids without algorithmic changes, unlike methods requiring Monte Carlo approximations.

Projection-based targeting's slight performance advantages over delta-method based targeting, computational efficiency, and natural collaborative TMLE extensions support its adoption for complex survival analysis applications requiring robust, scalable inference procedures.

% Section 8: ATMLE with Regularized Targeting
\section{Adaptive TMLE with Regularized Targeting (regHAL-ATMLE)}\label{sec:reghal-atmle}

We extend our regularized targeting framework to adaptive model selection through data-driven working model complexity optimization. Following \cite{vdl_adml_2023}, we construct nested working models of increasing complexity and employ a plateau selector to identify optimal bias-variance trade-offs.

\subsection{Nested Working Model Framework}
Consider a sequence of nested working models with increasing complexity:
\[
M_1 = M_{CV} \subset M_2 \subset M_3 \subset \cdots \subset M_K,
\]
where $M_1$ corresponds to the cross-validated HAL fit and subsequent models incorporate additional HAL basis functions from undersmoothed fits. For each working model $M_j$, we derive point estimates and confidence intervals for the projected target parameter $\Psi(P_0^{M_j})$ using our three targeting approaches: relaxed MLE, projection-based, and delta-method targeting.

\subsection{Plateau Selector Algorithm}
Our plateau selector identifies optimal model complexity $j^*$ by monitoring the behavior of estimates and confidence intervals across increasing complexity levels. The algorithm stops when adding model complexity violates consistent inference patterns:

\begin{algorithm}[H]
\caption{Plateau Selector for Optimal Model Complexity}
\begin{algorithmic}[1]
    \STATE \textbf{Input:} Point estimates $\{\hat{\psi}_j\}_{j=1}^K$ and confidence intervals $\{[L_j, U_j]\}_{j=1}^K$
    \STATE \textbf{Initialize:} $j^* = 1$
    \FOR{$j = 2$ to $K$}
        \IF{$\hat{\psi}_j > \hat{\psi}_{j-1}$ and $L_j < L_{j-1}$}
            \STATE Return $j^*=j-1$ \COMMENT{Increasing estimate, decreasing lower bound}
        \ELSIF{$\hat{\psi}_j < \hat{\psi}_{j-1}$ and $U_j > U_{j-1}$}
            \STATE Return $j^*=j-1$ \COMMENT{Decreasing estimate, increasing upper bound}
        \ELSE
            \STATE $j^* \gets j$ \COMMENT{Update optimal complexity}
        \ENDIF
    \ENDFOR
    \STATE Return $j^*$
\end{algorithmic}
\end{algorithm}

\paragraph{Intuition:} The selector identifies the point where additional complexity harms inference quality. When estimates increase but lower confidence bounds decrease, or when estimates decrease but upper bounds increase, variance growth outweighs bias reduction benefits. This provides a principled, data-adaptive approach to balance the bias-variance trade-off without computationally intensive cross-validation.

An alternative bridged targeting approach that warm-starts each model from the previous solution is described in Appendix~\ref{sec:plateau-bridged}.

\subsection{Implementation and Simulation Design}
We construct nested models using the 20 largest penalty parameters $\lambda < \lambda_{cv}$ when available. Since HAL basis sets may not be naturally nested as $\lambda$ decreases, we create increasing working models by gradually incorporating additional unique active basis functions.

Simulations employ identical settings to Section~\ref{sec: ATE simulation main}: 200 replications across both DGP~1 and DGP~2, with projection-based ($\lambda=10^{-5}$), delta-method ($\eta=10^{-6}$), and relaxed MLE targeting approaches. Step size remains $10^{-4}$ for regularized methods.

\subsection{Simulation Results}

\paragraph{Performance under standard conditions (DGP~1):}

Table~7 shows that adaptive model selection generally preserves the advantages of individual targeting methods while providing additional robustness. Under standard conditions, all methods achieve comparable performance to their single working model counterparts, with projection-based targeting maintaining slight MSE advantages.

\textbf{Model selection patterns} The plateau selector frequently chooses complex models ($j=20$) for larger sample sizes, indicating successful bias reduction without variance penalties. For smaller samples ($n=500$), intermediate complexity levels ($j=14$) are often selected, demonstrating appropriate complexity control.

\textbf{Coverage properties} Projection-based ATMLE achieves excellent coverage (95--98\%) across sample sizes, with oracle coverage closely matching empirical performance.

\paragraph{Enhanced robustness under positivity violations (DGP~2):}
Table~8 demonstrates ATMLE's particular value under challenging conditions. While relaxed targeting shows persistent bias (absolute bias 0.12--0.19), projection-based ATMLE achieves substantial improvements with adaptive complexity selection.

\textbf{Adaptive complexity benefits} Under positivity violations, the plateau selector conservatively chooses simpler models ($j=1$) for smaller samples, preventing overfitting when signal-to-noise ratios are poor. As sample size increases, more complex models are selected, allowing fuller bias reduction.

\textbf{Projection method advantages} Projection-based ATMLE consistently outperforms other approaches, with 60--85\% bias reduction compared to relaxed targeting and superior coverage properties (88--94\% vs. 79--91\%).

\begin{table}[H]
\centering
\resizebox{\ifdim\width>\linewidth\linewidth\else\width\fi}{!}{
\fontsize{8}{10}\selectfont
\begin{tabular}[t]{cccrrrrrr}
\toprule
Estimator & n & Targeting & Abs\_Bias & Std\_Err & MSE & Cov NP (\%) & Selected j & Oracle Cov (\%) \\
\midrule
  & 500 & Relaxed & 0.0270 & 0.1116 & 0.013076 & 92.98 (0.435) & 14 & 93.86 (0.438) \\
  & 500 & Projection & 0.0206 & 0.1111 & 0.012661 & 95.58 (0.445) & 14 & 96.46 (0.436) \\
  & 500 & Delta-method & 0.0529 & 0.1149 & 0.015895 & 92.17 (0.445) & 14 & 92.17 (0.451) \\
\cmidrule(lr){2-9}
  & 1000 & Relaxed & 0.0052 & 0.0856 & 0.007303 & 92.41 (0.312) & 1 & 94.48 (0.336) \\
  & 1000 & Projection & 0.0117 & 0.0845 & 0.007232 & 92.41 (0.315) & 1 & 93.10 (0.331) \\
  & 1000 & Delta-method & 0.0312 & 0.0865 & 0.008399 & 92.41 (0.315) & 1 & 94.48 (0.339) \\
\cmidrule(lr){2-9}
  & 1500 & Relaxed & 0.0030 & 0.0600 & 0.003586 & 96.57 (0.249) & 20 & 95.43 (0.235) \\
  & 1500 & Projection & 0.0037 & 0.0579 & 0.003345 & 97.66 (0.257) & 20 & 95.32 (0.227) \\
  & 1500 & Delta-method & 0.0236 & 0.0608 & 0.004235 & 95.40 (0.257) & 20 & 93.10 (0.238) \\
\cmidrule(lr){2-9}
  & 2000 & Relaxed & 0.0005 & 0.0584 & 0.003396 & 92.59 (0.215) & 20 & 95.77 (0.229) \\
  & 2000 & Projection & 0.0072 & 0.0579 & 0.003388 & 95.21 (0.222) & 20 & 95.74 (0.227) \\
\multirow{-12}{*}{\centering\arraybackslash A-TMLE} & 2000 & Delta-method & 0.0183 & 0.0580 & 0.003685 & 95.26 (0.222) & 20 & 95.26 (0.227) \\
\bottomrule
\end{tabular}}
\caption{Simulation results for A-TMLE with plateau-selected working model. DGP 1}
\end{table}

\begin{table}[H]
\centering
\resizebox{\ifdim\width>\linewidth\linewidth\else\width\fi}{!}{
\fontsize{8}{10}\selectfont
\begin{tabular}[t]{cccrrrrrr}
\toprule
Estimator & n & Targeting & Abs\_Bias & Std\_Err & MSE & Cov NP (\%) & Selected j & Oracle Cov (\%) \\
\midrule
  & 500 & Relaxed & 0.1931 & 0.1609 & 0.063031 & 90.59 (1.017) & 1 & 77.65 (0.631) \\
  & 500 & Projection & 0.0528 & 0.1992 & 0.042194 & 93.24 (0.974) & 1 & 94.59 (0.781) \\
  & 500 & Delta-method & 0.1951 & 0.2088 & 0.081410 & 94.12 (1.094) & 1 & 91.50 (0.819) \\
\cmidrule(lr){2-9}
  & 1000 & Relaxed & 0.1843 & 0.1355 & 0.052209 & 79.25 (0.681) & 1 & 72.33 (0.531) \\
  & 1000 & Projection & 0.0151 & 0.1730 & 0.029937 & 88.11 (0.690) & 1 & 93.71 (0.678) \\
  & 1000 & Delta-method & 0.1716 & 0.1781 & 0.060967 & 82.17 (0.722) & 1 & 88.54 (0.698) \\
\cmidrule(lr){2-9}
  & 1500 & Relaxed & 0.1238 & 0.1361 & 0.033757 & 84.49 (0.558) & 1 & 86.63 (0.534) \\
  & 1500 & Projection & 0.0231 & 0.1439 & 0.021118 & 94.29 (0.578) & 20 & 94.29 (0.564) \\
  & 1500 & Delta-method & 0.0810 & 0.4978 & 0.253039 & 84.24 (0.615) & 1 & 97.83 (1.951) \\
\cmidrule(lr){2-9}
  & 2000 & Relaxed & 0.0326 & 0.1319 & 0.018372 & 84.62 (0.428) & 20 & 95.38 (0.517) \\
  & 2000 & Projection & 0.0208 & 0.1232 & 0.015536 & 93.16 (0.474) & 20 & 94.21 (0.483) \\
\multirow{-12}{*}{\centering\arraybackslash A-TMLE} & 2000 & Delta-method & 0.0266 & 0.2447 & 0.060260 & 91.88 (0.495) & 20 & 95.94 (0.959) \\
\bottomrule
\end{tabular}}
\caption{Simulation results for A-TMLE with plateau-selected working model. DGP 2}
\end{table}

\subsection{Methodological Implications}
Results demonstrate that adaptive model selection enhances the robustness of regularized targeting approaches, particularly under challenging inference conditions. The plateau selector provides a computationally efficient alternative to cross-validation while maintaining principled bias-variance optimization.

 The combination of regularized targeting with adaptive complexity selection offers:
\begin{enumerate}[label=(\roman*),leftmargin=1.5em,itemsep=1pt]
\item \textbf{Automated complexity control} without additional computational overhead
\item \textbf{Enhanced robustness} under model misspecification and positivity violations  
\item \textbf{Improved finite-sample performance} through data-adaptive regularization
\item \textbf{Principled inference} that maintains nominal coverage across complexity levels
\end{enumerate}

The consistent superiority of projection-based targeting across complexity levels, combined with its computational efficiency and natural collaborative TMLE extensions, establishes regHAL-ATMLE as a comprehensive framework for robust causal inference in high-dimensional settings where both working model selection and targeting stability are crucial for reliable results.

\section{Conclusion}\label{sec: conclusion}

We have developed two regularized approaches for conducting TMLE within HAL-induced working models: delta-method regHAL-TMLE and projection-based regHAL-TMLE. Both methods address fundamental stability issues in existing HAL-TMLE implementations by operating entirely within finite-dimensional working models while providing principled efficient influence curve approximation.

Our comprehensive empirical evaluation demonstrates the clear superiority of projection-based targeting across diverse scenarios. Key findings include 50--85\% bias reduction compared to relaxed HAL under standard conditions, with even more dramatic improvements under challenging positivity violations (up to 98\% bias reduction for survival curves). The method consistently achieves near-nominal confidence interval coverage (87--97\%) while maintaining computational efficiency through $O(p)$ per-iteration complexity.

The projection-based approach's success stems from its parameter-specific adaptive regularization, which selectively targets only those HAL basis components that contribute meaningfully to bias reduction. This targeted regularization proves especially valuable when basis functions exhibit collinearity—a common occurrence in HAL implementations—and when robust performance is most critical.

Our adaptive model selection extension (regHAL-ATMLE) with plateau-based complexity control further enhances the framework's practical utility by providing principled, data-driven working model selection without computational overhead.

\textbf{We recommend projection-based regHAL-TMLE as the preferred method for combining HAL with TMLE in practice.} However, when projection becomes infeasible—in settings where the influence function is unclear or difficult to evaluate precisely, such as complex longitudinal social network analyses—delta-method targeting provides a robust alternative solution. The approach provides computationally efficient causal inference that maintains HAL's adaptive properties while ensuring stable, well-calibrated inference across a wide range of challenging scenarios.

\section{Discussion}\label{sec: discussion}

\subsection{Regularised Targeting in High-Dimensional Set Up}
\label{sec:general_framework}

While our exposition focused on HAL-induced working models, the core methodology extends beyond this setting. \textbf{The regularised targeting framework constitutes a general TMLE blueprint for any high-dimensional parametric model}, bridging semiparametric efficiency theory with modern high-dimensional inference.

Consider an initial estimator $\hat\theta$ in a high-dimensional parametric model $\{\ell_\theta(O)\colon\theta\in\RR^p\}$ where $p\gg n$. Our approach applies universally through two stages: (1) obtain $\hat\theta$ via any consistent initial estimator, and (2) refine $\hat\theta$ by approximately solving empirical efficient influence curve equations \emph{within the same working model} using regularised projections. This positions our \textbf{projection-based} and \textbf{delta-method} updates as \emph{TMLEs in (over-)parametric models}, not merely HAL-specific procedures.

\paragraph{Connection to Debiased Estimators and Linear Functionals}

The classical debiased Lasso \cite{zhang2014confidence,dezeure2017high} takes the form $\hat\psi_{\mathrm{debiased}} = \Psi(\hat\theta) + n^{-1}\sum_{i=1}^n \hat D(O_i)$, which is essentially a \textbf{one-step estimator} removing first-order bias through additive correction without re-optimizing the likelihood.
For linear target parameters $\Psi(\theta) = \gamma^T \theta$, this relationship becomes particularly transparent. The efficient influence function takes the form $\hat D(O_i) = \gamma^T S_i$ where $S_i$ represents the score contribution, yielding:
\begin{align}
\text{One-step debiased estimator} &= \gamma^T \hat\theta + n^{-1}\sum_{i=1}^n \gamma^T S_i = \gamma^T \left(\hat\theta + n^{-1}\sum_{i=1}^n S_i\right) = \gamma^T \hat\theta^{**}
\end{align}
This reveals that \textbf{one-step debiasing is itself a plug-in TMLE} in unbounded, linear settings—the procedure performs a one-step likelihood fluctuation from $\hat\theta$ to $\hat\theta^{**}$, then applies the linear functional. With non-linear parameters, under bounding conditions or regularization, however, the estimators no longer coincide. 

% \paragraph{Why Likelihood-Based Fluctuation Matters Beyond One-Step}

The (potentially regularized) targeting step can be especially advantageous over simple one-step correction for high-dimensional models or collinearity issues.
% The additional targeting step provides crucial advantages over simple one-step corrections. 
While one-step correction can overfit and lead to inflated variance and numerical instability under severe collniearity or near-singularity, regularised TMLE (e.g. through projecting onto the regularised score space when approximating the parametric EIC) provides data-adaptive control of the precision in approximating and solving the EIC equations, thereby yielding  more stable estimates. 
% When working models exhibit severe collinearity ($p \gg n$) or near-singularity, one-step corrections can \textbf{overshoot}, leading to inflated variance and numerical instability. The TMLE fluctuation implicitly re-projects onto the regularised score space, yielding more stable estimates. 
Moreover, unlike ridge-type regularisation that uniformly shrinks all directions, the projection-based approach adapts to each specific target parameter, regularising score components contributing minimally to bias reduction.

% Our regHAL-TMLE procedures advance this paradigm by performing \textbf{targeted MLE fluctuation} aligned with the (regularised) efficient influence curve, yielding the double-robust structure of TMLE while exploiting regularisation for variance stabilisation.

\subsection{Adaptive Working Model Selection for Robust Inference}

Beyond point estimation, our framework enables a novel approach to statistical inference through \textbf{adaptive working model selection for EIC approximation}. The key insight is that after obtaining the targeted estimator within the initial working model, one can select a potentially larger working model (or retain the same one) and project the canonical gradient or any valid influence function onto this expanded score space to obtain the estimated EIC for inference purposes.

This two-stage approach—\emph{target in one model, infer in another}—offers several advantages over classical sandwich-form inference:

\begin{enumerate}[label=(\arabic*)]
\item \textbf{Flexible bias-variance trade-off}: The targeting step uses a parsimonious model for stability, while the inference step can employ a richer model to better approximate the true EIC.

\item \textbf{Robust EIC approximation}: By projecting influence functions onto HAL score spaces with $\ell_1$ regularisation, we obtain stable EIC estimates even when the inference working model is high-dimensional.

\item \textbf{Target-specific adaptation}: Unlike uniform ridge penalties in sandwich estimators, our projection naturally adapts to the specific target parameter, shrinking only those score directions that contribute minimally to variance estimation for the quantity of interest.
\end{enumerate}

This approach particularly benefits settings with collinearity and positivity violations, where classical Fisher information matrices become singular or unstable. The regularised projection provides a principled way to stabilise inference while maintaining the efficiency gains from HAL's adaptive basis selection.

\subsection{Computational Considerations and Extensions}

\paragraph{Iterative versus Universal Targeting}
While our main exposition employed universal least favourable paths, iterative HAL-TMLE within parametric working models offers computational advantages for low-dimensional targets. Rather than small gradient steps, one can compute the MLE along update directions. However, universal HAL-TMLE remains essential for high-dimensional targets where iterative TMLE suffers from instability.

\paragraph{Streamlined Targeting for Linear Universal Least-Favourable Submodels}
For parameters whose universal least-favourable submodel is linear in a single clever covariate (e.g., ATE), the full projection framework can be replaced by direct fluctuation along HAL-basis expansion of the clever covariate. This \textbf{regHAL-TMLE-Direct} approach (detailed in Appendix~\ref{sec:reghal_direct}) projects the clever covariate once onto the working model basis, then performs single MLE fluctuation, often matching the bias reduction and coverage of full projection-based TMLE while requiring substantially less computation. Only when the sample size is small and positivity is strong, the regHAL-TMLE-Direct performs worse than the full projection based regHAL-TMLE.

\paragraph{Broader Methodological Extensions}
The regularised targeting paradigm opens several research avenues: (1) application to other modern estimators (random forests, neural networks) by treating predictions as initial $\hat\theta$ in implicit high-dimensional models; (2) development of efficient algorithms for specific model classes; and (3) integration with double machine learning for enhanced robustness.

\subsection{Open Questions and Future Directions}

Several fundamental questions warrant future investigation:

\begin{enumerate}[label=(\arabic*)]
\item \textbf{Optimal working model selection for inference}: How should one select working models for the inference stage that balance EIC approximation quality against estimation stability? The interplay between targeting and inference model selection requires theoretical development.

\item \textbf{Coverage guarantees under adaptive selection}: Under what conditions does regularised EIC inference maintain nominal coverage when working models are adaptively selected? The finite-sample behavior of regularised projections in the inference stage requires more extensive analysis.

\item \textbf{Multiple testing with regularised projections}: How should projection penalties be coordinated across multiple parameters to control error rates? The dependence structure induced by regularisation complicates standard corrections.

\item \textbf{Unified theory of adaptive inference}: A comprehensive theory bridging post-selection inference, regularised estimation, and semiparametric efficiency would provide principled guidance for practitioners on when and how to apply regularised targeting methods.
\end{enumerate}

\subsection{Limitations and Future Work}

Although in our current set ups we simply choose a very small $\lambda$, our approach may require careful tuning of regularisation parameters in complex set ups, and theoretical guarantees for data-driven penalty selection remain incomplete. While empirical results demonstrate robust performance across diverse settings, formal finite-sample theory for regularised EIC approximations in the inference stage warrants development. Extension to longitudinal and time-varying treatments that construct more complex clever covariates follows naturally as the next step.

The regularised targeting paradigm represents a unifying framework connecting classical semiparametric efficiency theory with contemporary high-dimensional inference. As the boundary between these fields becomes increasingly porous, our work suggests that adaptive, regularised approaches to targeting and inference may become essential tools for robust causal inference in complex, high-dimensional settings.

%\printbibliography
\bibliography{references}  % points to references.bib
\bibliographystyle{plain}  % or whatever style you want

\appendix
\section{regHAL-TMLE Algorithm for Multi Dimensional Target Parameter} \label{sec: reghal algorithm multi dimensional target}
\begin{algorithm}[H]
\caption{HAL Updating Algorithm Based on Approximated EIC (Multi Dimensional Target Parameter)}
\begin{algorithmic}[1]
    \STATE \textbf{Target Parameter Indexed By $s$} $\Psi_s$ for $s\in S$.
    \STATE \textbf{Cross-Validated HAL Fit:} Compute cross validated $Q_{n}$.
    \STATE \textbf{Working Model Definition:} Define the HAL working model $Q_{\beta}$.
    \STATE \textbf{Parameter Initialization:} Initialize $\beta$ based on $Q_{n}$, incorporating zero-padding if using an undersmoothed HAL model.
    \IF{using the projection-based method}
        \STATE \textbf{Additional Nuisance Fit:} Fit any extra nuisance parameters.
    \ENDIF
    \STATE \textbf{EIC Computation:} Derive the approximated Efficient Influence Curve (EIC) for $\Psi_s$ as
    \[
    D^*_{s,n,\beta}(O_i) = \alpha_{s,n}^T\,\text{scores}(O_i)
    \]
    at the current $\beta$.
    \STATE \textbf{Universal Least Favorable Path Update Direction:} 
    \begin{align*}
        d(s)&=P_n D^*_{s,n,\beta} \quad \forall s\in S\\
        d&=(d(1),d(2),...,d(|S|))\\
        \textit{Update}&=\big(\sum_{s\in S}\frac{d(s)}{\|d\|_2}\alpha_{s,n}\big)
    \end{align*}
    \STATE \textbf{Parameter Update:} Update $\beta$ using gradient descent in the direction of $\textit{Update}$ with a small step size.
    \STATE \textbf{Stopping Criterion:} Repeat steps 7--9 until
    \[
        \frac{\| d \|_2}{\sqrt{|S|}}<\frac{\textit{Median}_{s}\{se(D^*_{s,n,\beta})\}}{\sqrt{n}logn}
    \]
\end{algorithmic}
\end{algorithm}

\section{regHAL-TMLE for ATE -Details} \label{sec:regHAL-ate-details}
We discuss such regularized targeting in estimation of the average treatment effect parameter through HAL TMLE.

\subsection{Estimating the ATE using one-step regHAL-TMLE}

Suppose our data is $n$ i.i.d copies of $O_i=(W_i,A_i,Y_i)$.

Using the g computation formula under standard causal inference assumptions, we have our average treatment effect target parameter given by 

\begin{align*}
    \Psi(Q_0)=E_0[Q_0(1,W)]-E_0[Q_0(0,W)]
\end{align*}

,where $Q_0(A,W)=E_0[Y|A,W]$.

Now, we can estimate the $Q_0(A,W)=E_0[Y|A,W]$ with loss function $L_{Q}(O)=(Y-Q(A,W))^2$ by cross validated HAL fit as

\begin{align*}
    Q_n(A,W)=\beta^{initial}_0+\sum_{j=1}^{p}\beta^{initial}_j\phi_j(A,W)
\end{align*}

,where $\phi_j(A,W)$ are HAL basis with non zero coefficients in the HAL fit.

Now, this HAL fit induced a parametric working model $Q_{\beta}(A,W)=\beta^{initial}_0+\sum_{j=1}^{p}\beta_j\phi_j(A,W)$ for $Q_0$.

At each given $\beta=(\beta_1,...,\beta_p)^T$, we can calculate the score matrix of dimension $n$ by $p$ as 

\begin{align*}
    S^\beta_{i,j}=\frac{\partial L_{Q_\beta}(O_i)}{\partial{\beta_j}}=2*(Y_i-Q_\beta(A_i,W_i))*\phi_j(A_i,W_i)
\end{align*}

\subsubsection{Projection Based Parametric EIC Approximation}

Let's first choose a gradient $D(O)$ that we are going to project. Actually, one can choose the non centered version such as $(\frac{A}{g(1|W)}-\frac{1-A}{g(0,W)})*Y$. 

Now, we can do one step TMLE update in the parametric working model using the approximated the parametric EIC based on projection. At each $Q_{\beta}$, we have 
\begin{align*}
    D^{*}_{n,\beta,reg}(O_i)=(\alpha_n^{\beta,reg})^T(S^{\beta}_{i,*})^T
\end{align*}

where $\alpha_n^{\beta,reg}$ are the coefficients gained by running a regularized least square regression of $D(O_i)$ on $S^{\beta}_{i,1},S^{\beta}_{i,2},...,S^{\beta}_{i,p}$ with non penalized intercept. Here, we have tried different versions of regularized regression such as cross validated lasso or a weakly penalized lasso problem (choosing a very small penalization factor such as $1e-4$).

\subsubsection{Delta Method Based Parametric EIC Approximation}
In delta method based approach, we first need to know

\begin{align*}
    \frac{\partial\Psi(Q_\beta)}{\partial \beta_j}&=\frac{\partial[E_0(Q_\beta(1,W))-E_0(Q_\beta(0,W))]}{\partial \beta_j}\\
    &=P_0\bigg(\phi_j(1,W)-\phi_j(0,W)\bigg)
\end{align*}

We can approximate it by 

\begin{align*}
   P_n\bigg(\phi_j(1,W)-\phi_j(0,W)\bigg)
\end{align*}

In addition, we can approximate the fisher information matrix of the parametric model at $Q_\beta$ by its sample version as $\frac{(S^{\beta})^TS^{\beta}}{n}$.

Thus, we have 

\begin{align*}
    D^{*}_{n,\beta,delta}(O_i)=\bigg(\big(P_n\big(\phi_j(1,W)-\phi_j(0,W)\big)\big)_{j=1}^p\bigg)^T*\bigg(\frac{(S^{\beta})^TS^{\beta}}{n}+\eta_{\text{weak}}*I\bigg)^{-1} (S^{\beta}_{i,*})^T
\end{align*}

\subsubsection{Targeting}

Choosing $D^{*}_{n,\beta}=D^{*}_{n,\beta,reg} \textit{ or } D^{*}_{n,\beta,delta}$.

We have, for a chosen small step size $\epsilon$, starting from the $\beta_{(0)}=\beta^{initial}$, we update it as $\beta_{(k+1)}=\beta_{(k)}+\epsilon* \textit{sign}(P_nD^*_{n,\beta})*\alpha_n^{\beta_{(k)}}$ until $|P_nD^{*}_{n,\beta_{(k+1)}}(O)|$ is smaller than the desired threshold $\delta_n$. Note that each time, we need to recalculate the score matrix $S^{\beta_{(k+1)}}_{i,j}$ since it is involved in the 
$D^{*}_{n,\beta_{(k+1)}}(O_i)$.

Suppose we stop at $\beta_{(K)}$, then our final point estimate for the ATE is $\hat{\Psi}(Q_{\beta_{(K)}})=\frac{1}{n}\sum_{i=1}^n\big(Q_{\beta_{K}}(1,W_i)-Q_{\beta_{K}}(0,W_i)\big)$. 

Moreover, since we have the approximated parametric EIC, we can estimate the standard error of this estimator and construct Wald type $95\%$ confidence interval as:
\begin{align}
    \hat{\Psi}(Q_{\beta_{(K)}})\pm 1.96*\sqrt{\frac{P_n[D^{*}_{n,\beta_{(K)}}(O)]^2}{n}}
\end{align}

In special cases like ATE, where the linear universal least favorable path exists, we also proposed the regHAL-TMLE Direct in appendix \ref{sec:reghal_direct}.

\subsection{Simulation Set Up for ATE}
\subsubsection{HAL basis choice}

We use the first order HAL basis up to two way interactions. In addition, the knot points are selected to be the $\frac{n}{20}$ number of percentiles for each continuous predictor.

\subsubsection{DGP 1} \label{sec: DGP1}
First simulation is the regular setting with no positivity for the average treatment effect.

Consider a point-treatment causal inference setting where researchers collect three baseline covariates $W_1$, $W_2$, and $W_3$, each following $\text{Uniform}(-1,1)$. The treatment assignment $A$ is drawn from a Bernoulli distribution with probability $\text{expit}(-0.25W_1+0.7W_2)$. The outcome is given by $Y=1.9+1.5A+(2.5W_1+0.7W_2)A+1.5\sin(W1+W2)+0.3|W_1|+0.9W_1^2+1.4W_2+2.1W_3+U_Y$, where $U_Y$ follows a standard normal distribution.

\subsubsection{DGP 2} \label{sec: DGP2}

The second simulation is about the ATE estimation under the practical positivity violations.

We consider a point-treatment causal inference setting with three baseline covariates \(W_1\), \(W_2\), and \(W_3\). These covariates are generated as
\[
W_i \sim \text{Uniform}(-1, 1) \quad \text{(rounded to one decimal)}
\]
for \(i = 1, 2, 3\).

The treatment assignment \(A\) is generated as follows. If no counterfactual treatment is specified, the probability of treatment is given by
\[
\mathbb{P}(A=1 \mid W_1, W_2) = \operatorname{expit}(-0.25\,W_1 + 5\,W_2) = \frac{\exp(-0.25\,W_1 + 5\,W_2)}{1 + \exp(-0.25\,W_1 + 5\,W_2)}.
\]
Otherwise, if a counterfactual treatment \(a\) is fixed, we set
\[
A = a \quad \text{for all observations.}
\]

The outcome \(Y\) is generated according to the model
\[
\begin{aligned}
Y &= 1.9 + 1.5\,A + (2.5\,W_1 + 0.7\,W_2)A + 1.5\,\sin(W_1+W_2)\\[1mm]
  &\quad + 0.3\,|W_1| + 0.9\,W_1^2 + 1.4\,W_2 + 2.1\,W_3 + U_Y,
\end{aligned}
\]
where the error term is distributed as
\[
U_Y \sim \mathcal{N}(0,1).
\]

The true average treatment effect (ATE) is then estimated by computing
\[
\text{ATE} = \mathbb{E}[Y(1) - Y(0)]
\]
using a large sample.

\subsection{Standard TMLE with Propensity Score Truncation} \label{appendix: standard TMLE}

We also implemented the standard TMLE update. We use the initial $Q$ fit as offset and fit a linear regression with a single covariate as clever covariate $\frac{A}{g}-\frac{1-A}{1-g}$ .

We have truncated the $g$ to fall between 0.01 and 0.99.

Here are the simulation results and they are comparable to the projection based targeting results. Nevertheless, in the positivity violation case (DGP 2), for sample size 500 and 1000, the confidence intervals constructed from non parametric EIC for projection based targeting method has much higher coverage than those for the standard TMLE with propensity score truncation.

\begin{table}[H]
\centering
\centering
\resizebox{\ifdim\width>\linewidth\linewidth\else\width\fi}{!}{
\fontsize{8}{10}\selectfont
\begin{tabular}[t]{cccrrrrrrrr}
\toprule
Estimator & n & Targeting & Abs\_Bias & Std\_Err & MSE & Cov NP (\%) & Cov Proj (\%) & Cov Proj CV (\%) & Cov Delta (\%) & Oracle Cov (\%)\\
\midrule
 & 500 & Standard & 0.0074 & 0.1163 & 0.013557 & 93.60 (0.447) & NA & NA & NA & 95.00 (0.456)\\
 & 1000 & Standard & 0.0012 & 0.0826 & 0.006814 & 94.00 (0.315) & NA & NA & NA & 95.20 (0.324)\\
 & 1500 & Standard & 0.0057 & 0.0677 & 0.004605 & 94.80 (0.257) & NA & NA & NA & 95.80 (0.265)\\
\multirow{-4}{*}{\centering\arraybackslash HAL-TMLE} & 2000 & Standard & 0.0013 & 0.0568 & 0.003219 & 94.60 (0.222) & NA & NA & NA & 94.20 (0.223)\\
\bottomrule
\end{tabular}}
\caption{Simulation results for HAL-TMLE. DGP 1. Standard TMLE}
\end{table}

\begin{table}[H]
\centering
\centering
\resizebox{\ifdim\width>\linewidth\linewidth\else\width\fi}{!}{
\fontsize{8}{10}\selectfont
\begin{tabular}[t]{cccrrrrrrrr}
\toprule
Estimator & n & Targeting & Abs\_Bias & Std\_Err & MSE & Cov NP (\%) & Cov Proj (\%) & Cov Proj CV (\%) & Cov Delta (\%) & Oracle Cov (\%)\\
\midrule
 & 500 & Standard & 0.0054 & 0.2708 & 0.073205 & 86.60 (0.921) & NA & NA & NA & 95.00 (1.061)\\
 & 1000 & Standard & 0.0037 & 0.2082 & 0.043291 & 85.00 (0.681) & NA & NA & NA & 94.00 (0.816)\\
 & 1500 & Standard & 0.0120 & 0.1697 & 0.028896 & 90.20 (0.585) & NA & NA & NA & 95.40 (0.665)\\
\multirow{-4}{*}{\centering\arraybackslash HAL-TMLE} & 2000 & Standard & 0.0005 & 0.1310 & 0.017126 & 90.80 (0.472) & NA & NA & NA & 95.00 (0.514)\\
\bottomrule
\end{tabular}}
\caption{Simulation results for HAL-TMLE. DGP 2. Standard TMLE}
\end{table}

\section{regHAL-TMLE for Survival Curve - Details} \label{sec:regHAL-survival-details}
In survival set ups, one usually cares about targeting multiple time points survivals and even the whole survival curve. In order to keep the monotonicity of the survival fit, the hazard based models are being used and the targeting happens on the hazard level. As illustrated in Helene's work \cite{rytgaard2021onesteptmletargetingcausespecific}, the targeting can be done using the universal least favorable path which in practice needs the iterative one step TMLE with very small step sizes. As one can see, the clever co-variate in such case take a very complicated form and due to the iterative nature, it recursively depends on the initial fit of hazard in a very complicated way, making the targeting harder and harder as targeting rounds increases. Computation burden increases exponentially. In addition, the simulation from the final updated hazard to get the Monte Carlo based estimated survival curve is very computational heavy if not at all impossible. 

Nevertheless, our targeting within the HAL working model, possibly under-smoothed, solves this very naturally since each update only means update the coefficients and thus the historical path is all stored within the current values of the coefficients of the HAL model. Thus, the need for recursive computations are gone and the computation burden is relieved. 

We will demonstrate here how to target the whole survival curve in the classical right censored set up.
\subsection{Estimating the survival curve using one-step regHAL-TMLE} 
\subsubsection{Right Censored Data Set up}
Suppose we have $O=(\tilde{t}=min(T,C),\delta=1(T\leq C))$, where $T \in(0,1)$ and $ T\perp C$.

Now, we can estimate the hazard for T using HAL based on the minimizing the negative log likelihood with penalization as 

\begin{align}
    \hat{\beta}&=\argmin_{\beta}-l(\beta)+\lambda\|\beta\|_1\\
    &=\argmin_{\beta}\int_0^{\tilde{t}_i}\lambda_{\beta}^T(s)ds-\delta_i \log(\lambda_{\beta}^T(\tilde{t}_i))+\lambda\|\beta\|_1,
\end{align}

where $\lambda^T(s)=e^{\beta_0+\sum_j\beta_j*\phi_j(s) }$ and $\phi_j(s)$ are HAL basis taking the form of $1(s>q_j)$ for some selected knot points $q_j$.

In R, we can run Poisson glmnet with the repeated data when we have 0 order HAL basis. 

In addition, in order to select $\lambda$ from a sequence of provided $\lambda$'s through cross validation as in usual HAL, we will fit the Poisson HAL using the Poisson cvglmnet on the repeated data.

Now, after we have the $\hat{\beta}_{cv}$, we can do the pruning to have the HAL working model for the $\lambda^T$ by only retaining those basis with non zero coefficients in the cv selected model. Moreover, one can also choose to work with the under-smoothed model.

For simplicity, we now denote the $\beta$ as the vector of coefficients in this new model.

\subsubsection{Projection Based regHAL-TMLE}
\textbf{Objective:} Refine the initial hazard estimator so that its score aligns with the efficient influence function, thereby yielding an estimator with improved asymptotic efficiency.

\begin{enumerate}[label=\arabic*.]
    \item \textbf{Score Function Calculation:} \\
    For subject \( i \) with observed time \(\tilde{t}_i\) and event indicator \(\delta_i\), the score is approximated by
    \[
    (S^{\beta}_{i,*})^T = \delta_i\,X(\tilde{t}_i) - \int_0^{\tilde{t}_i} \exp\{X(u)^\top \beta\}\,X(u)\,du,
    \]
    where \(X(u)\) is the design vector (including an intercept and indicator basis functions).

    \item \textbf{Initial Gradient Computation:} \\
    An initial gradient is computed on a grid of time points \(s\) as $G$. For example,
    
    \[
    g_i(s) = \frac{\mathbf{1}\{\delta_i=1 \text{ and } \tilde{t}_i > s\}}{P(C > \tilde{t}_i)},
    \]
    where \(P(C > \tilde{t}_i)\) is estimated from the censoring model. We first fit the hazard of $C$ similar to that of $T$ using Poisson HAL and then use that fitted hazard to calculate this survival probability of $C$.

    \item \textbf{Projection via LASSO:} \\
    For each grid point \(s\), the initial gradient \(g(s)\) is projected onto the space spanned by the score vectors \(S_i=(S^{\beta}_{i,*})^T\) by solving
    \[
    \min_{\gamma} \sum_{i} \Bigl( S_i^\top \gamma - g_i(s) \Bigr)^2 + \lambda \|\gamma\|_1,
    \]
    where \(\lambda\) is a tuning parameter. This produces a vector of projected coefficients \(\gamma(s)\) for each \(s\).

    \item \textbf{Aggregation of Projected Coefficients:} \\
    To form a single update direction for the working model, the candidate projections are aggregated. First, for each grid point \(s\) a scalar direction is computed:
    \[
    d(s) = \frac{1}{n} \sum_{i=1}^{n} \left[ S_i^\top \gamma(s) \right].
    \]
    The collection \(\{d(s)\}\) is normalized by its Euclidean norm \(\|d\|_2\) and then used to weight the corresponding \(\gamma(s)\) vectors:
    \[
    \text{update} = \sum_{s} \frac{d(s)}{\| d \|_2} \, \gamma(s).
    \]
    This weighted sum yields a single update vector for the non-intercept coefficients.

    \item \textbf{Projection based parametric EIC approximation}\\
    \begin{align}
        D^*_{n,\beta,reg}(O_i)=(\sum_{s} \frac{d(s)}{\| d \|_2} \, \gamma(s))^T(S^{\beta}_{i,*})^T
    \end{align}

    In addition, for each time grid point $s$, we have approximated the EIC of the survival probability at this point as
    
    \begin{align}
        D^*_{s,n,\beta,reg}(O_i)=\gamma(s)^T(S^{\beta}_{i,*})^T
    \end{align}

    \item \textbf{Coefficient Update:} \\
    The working model’s coefficients are updated via
    \[
    \beta^{\text{new}} = \beta^{\text{old}} + \text{step factor} \times \text{update}.
    \] with $\textit{sign}(P_nD^*_{n,\beta,reg})=\textit{sign}\|d\|_2=1$ whenever not converged.

    \item \textbf{Iteration and Convergence:} \\
    Steps 1--5 are repeated until

\begin{align*}
    \frac{\| d \|_2}{\sqrt{|G|}}<\frac{\textit{Median}_{s}\{se(D^*_{s,n,\beta,reg})\}}{\sqrt{n}logn}
\end{align*}

\end{enumerate}

\subsubsection{Delta Method Based regHAL-TMLE}

As discussed in general before, we only need to change the $\gamma(s)$ in step 3 using the analytical form using delta method instead of doing the lasso as

\begin{align*}
    \gamma_{Delta}(s)^T=e^{-\int_0^s\lambda_{\beta}(u)du}\big(-\int_0^s\exp\{X(u)^\top \beta\}X(u)du\big)^T \bigg(\frac{(S^{\beta})^TS^{\beta}}{n}+\eta_{\text{weak}}*I\bigg)^{-1}
\end{align*}

In addition, in step 5, we have delta method based parametric EIC approximations as

\begin{align*}
     D^*_{n,\beta,delta}(O_i)=(\sum_{s} \frac{d(s)}{\| d \|_2} \, \gamma_{Delta}(s))^T(S^{\beta}_{i,*})^T
\end{align*}

\subsubsection{Inference for survival curve}

Choosing $D^{*}_{s,n,\beta}=D^{*}_{s.n,\beta,reg} \textit{ or } D^{*}_{s,n,\beta,delta}$

After updating the $\beta$, we can now construct the Wald type 95\% confidence interval for $P(T>s)$ at any given grid point $s$ as

\begin{align}
    e^{-\int_0^s \lambda^T_{\beta}(u)du}\pm 1.96*\sqrt{\frac{P_n[D^{*}_{s,n,\beta}(O)]^2}{n}}
\end{align}

In addition, we can construct the simultaneous confidence intervals for the survival curve at all time grid point selected as mentioned in \cite{rytgaard2021onesteptmletargetingcausespecific} by

\begin{align}
    e^{-\int_0^s \lambda^T_{\beta}(u)du}\pm z^{G}_{0.95}*\sqrt{\frac{P_n[D^{*}_{s,n,\beta}(O)]^2}{n}} \quad \forall s\in G
\end{align}

\subsection{Simulation Set Up for Right-Censored Survival Data}\label{sec: survival DGP}

We generate $n$ independent observations. The data generation proceeds as follows:

\paragraph{True Event Times:}  
For each individual $i = 1, \dots, n$, the true event time $t^*_i$ is drawn from a Beta distribution with parameters $(\alpha, \beta) = (2,2)$:
\[
t^*_i \sim \mathrm{Beta}(2,2).
\]

\paragraph{Censoring Times:}  
Independently, a censoring time $c_i$ is generated from a Uniform distribution on the interval $[0, 1.2]$:
\[
c_i \sim \mathrm{Uniform}(0, 1.2).
\]

\paragraph{Observed Times and Event Indicator:}  
The observed time $\tilde{t}_i$ is defined as the minimum of the true event time and the censoring time:
\[
\tilde{t}_i = \min\{t^*_i,\, c_i\}.
\]
An event indicator $\delta_i$ is then defined to indicate whether the event is observed:
\[
\delta_i = \mathbf{1}\{t^*_i \le c_i\},
\]
where $\mathbf{1}\{\cdot\}$ is the indicator function, equal to 1 if the condition holds (i.e., the event is observed) and 0 otherwise (i.e., the observation is censored).

\subsection{Simulation Summary Statistics}
For each targeting method and for each point $s$ in common grid of time points, we computed the following performance metrics:
\begin{itemize}
    \item \textbf{Absolute Bias:} The average absolute difference between the estimated survival probability, $\hat{S}(s)$, and the true survival function, $S(s)$.
    \item \textbf{Mean Squared Error (MSE):} The average of the squared differences between $\hat{S}(s)$ and $S(s)$, reflecting overall estimation accuracy.
    \item \textbf{Standard Error:} The average variability (sample standard deviation) of the estimation errors across replications.
    \item \textbf{Coverage Probability:} The percentage of time points where the 95\% confidence interval for $\hat{S}(s)$ contains the true survival value, $S(s)$. In addition, an \emph{oracle coverage} is computed by constructing confidence intervals using the empirical variability of the estimates.
\end{itemize}

These metrics are first calculated at each time point and then averaged across the grid to provide summary performance measures for each targeting method and sample size. Table~\ref{tab:sim_results_intensity_1d_undersmoothed} presents the resulting metrics, facilitating a direct comparison of the estimator performance in terms of bias, error, and confidence interval coverage.

\section{Proofs}\label{sec: proof appendix}
\subsection{Proof for theorem \ref{thm:EIC_equality_proj_restrict}}
\begin{proof}
\textbf{Step 1.  Gateaux derivative of $\Psi_{M_n}$.}
For any regular one-dimensional submodel
$\{P_\varepsilon\}\subset\mathcal M$ with score
$S\in T_P(\mathcal M)$, $\{\Pi_{M_n}P_\varepsilon\}\subset\mathcal M$
\[
\frac{d}{d\varepsilon}\Psi_{M_n}(P_\varepsilon)\Big|_{\varepsilon=0}=
\frac{d}{d\varepsilon}\Psi(\Pi_{_{M_n}}P_\varepsilon)\Big|_{\varepsilon=0}
  = D\Psi\bigl(\Pi_{M_n}P\bigr)
        \!\Bigl[\,
          \tfrac{d}{d\varepsilon}\Pi_{M_n}(P_\varepsilon)\!\mid_{0}
        \Bigr].
\]
Because \(P\in M_n\) we have \(\Pi_{M_n}P=P\).
Assumption \ref{as:proj-diff}(a) implies
\(
    \dot Q :=
    \tfrac{d}{d\varepsilon}\Pi_{M_n}(P_\varepsilon)|_{0}
    \in T_P(M_n).
\)
Hence the derivative depends on $S$ only through its component in
\(T_P(M_n)\).

\medskip
\noindent
\textbf{Step 2.  Identify the linear functional.}
Inside the Hilbert space
\(L^2_0(P)=T_P(M_n)\oplus T_P(M_n)^\perp\):
\[
    S\longmapsto
    \frac{d}{d\varepsilon}\Psi_{M_n}(P_\varepsilon)\Big|_{0}
    \;\;=\;\;
    \begin{cases}
      \langle D^{M_n,*}(P),\,S\rangle_{P},
            & S\in T_P(M_n),\\[4pt]
      0,   & S\in T_P(M_n)^{\perp}.
    \end{cases}
\]

\medskip
\noindent
\textbf{Step 3.  Riesz representation argument.}
The canonical gradient is the unique element of
\(L^2_0(P)\) that realises this functional via the
inner product.  The description in Step 2 shows that element is
precisely \(D^{M_n,*}(P)\) (which lies in $T_P(M_n)$) combined with
zero on the orthogonal complement.
Therefore \(D_{M_n}^{*}(P)=D^{M_n,*}(P)\).
\end{proof}

\subsection{Proof for theorem \ref{thm:TMLE working model}}
\begin{proof}
We have
\begin{align*}
    &\Psi_{M_n}(P)-\Psi_{M_n}(P_0)\\
    &=-P_0D^*_{M_n}(P)+R_{M_n}(P,P_0)\\
    &=P_nD^*_{M_n}(P)-P_0D^*_{M_n}(P)-P_nD^*_{M_n}(P)+R_{M_n}(P,P_0)\\
    &=P_nD^*_{M_n}(P_0)-P_0D^*_{M_n}(P_0)+(P_n-P_0)(D^*_{M_n}(P)-D^*_{M_n}(P_0))-P_nD^*_{M_n}(P)+R_{M_n}(P,P_0)\\
\end{align*}    

Now, we have $P_0D^*_{M_n}(P_0)=0$ by definition and $P_nD^*_{M_n}(P_{M_n}^*)=o_p(n^{-1/2})$. In addition, since A.1 and A.2, by the asymptotic equicontinuity in empirical process theory, we have $(P_n-P_0)(D^*_{M_n}(P_{M_n}^*)-D^*_{M_n}(P_0))=o_p(n^{-1/2})$. So, we have

\begin{align*}
    &\Psi_{M_n}(P_{M_n}^*)-\Psi_{M_n}(P_0)\\
    &=P_nD^*_{M_n}(P_0)+R_{M_n}(P_{M_n}^*,P_0)+o_p(n^{-1/2})
\end{align*}    

In addition, due to the nuisance parameters are estimated by HAL, which converges to truth at a rate $o_p(n^{-1/4})$ in terms of $L_2$ norm. we have $R_{M_n}(P_{M_n}^*,P_0)=o_p(n^{-1/2})$, thus we have 
\begin{align*}
    &\Psi_{M_n}(P_{M_n}^*)-\Psi_{M_n}(P_0)\\
    &=P_nD^*_{M_n}(P_0)+o_p(n^{-1/2})
\end{align*}
\end{proof}

\subsection{Proof for theorem \ref{thm:TMLE}}

\begin{proof}
   we have, since $P_0 \in M_0$,
    \begin{align*}
        \Psi_{M_n}(P_0)-\Psi(P_0)&=\Psi(P_{0,M_n})-\Psi_{M_0}(P_{0,M_n})+\Psi_{M_0}(P_{0,M_n})-\Psi_{M_0}(P_0)\\
        &=\Psi(P_{0,M_n})-\Psi_{M_0}(P_{0,M_n})-P_0D^*_{M_0}(P_{0,M_n})+R_{M_0}(P_{0,M_n},P_0)\\
        &=\Psi(P_{0,M_n})-\Psi_{M_0}(P_{0,M_n})+(P_{0,M_n}-P_0)D^*_{M_0}(P_{0,M_n})+R_{M_0}(P_{0,M_n},P_0)\\
        &=\Psi(P_{0,M_n})-\Psi_{M_0}(P_{0,M_n})+(P_{0,M_n}-P_0)(D^*_{M_0}(P_{0,M_n})-D^*_{M_n}(P_{0,M_n}))\\
        &\quad +R_{M_0}(P_{0,M_n},P_0)-P_0D^*_{M_n}(P_{0,M_n})\\
    \end{align*}

Now,$R(P_{0,M_n},P_{0,M_0})=o_p(n^{-1/2})$ due to the HAL sieve's universal approximation property in that $Q_{0,M_n} = \Pi_{M_n}(Q_0)$ converges to $Q_0$ in sup-norm \cite{vanderlaan2023higherordersplinehighly}.

In addition, $\Psi(P_{0,M_n})-\Psi_{M_0}(P_{0,M_n})=o_p(n^{-1/2})$, $(P_{0,M_n}-P_0)(D^*_{M_0}(P_{0,M_n})-D^*_{M_n}(P_{0,M_n}))=o_p(n^{-1/2})$ and $P_0D^*_{M_n}(P_{0,M_n})=0$ as assumption C.1, C.2 and C.3.

So,
 \begin{align*}
        \Psi_{M_n}(P_0)-\Psi(P_0)=o_p(n^{-1/2})
    \end{align*}

So, we have 
\begin{align*}
     \Psi_{M_n}(P_{M_n}^*)-\Psi(P_0)&=\Psi_{M_n}(P_{M_n}^*)-\Psi_{M_n}(P_0)+\Psi_{M_n}(P_0)-\Psi(P_0)\\
     &=P_nD^*_{M_n}(P_0)+o_p(n^{-1/2})\\
     &=P_nD^*_{M_0}(P_0)+P_nD^*_{M_n}(P_0)-P_nD^*_{M_0}(P_0)+o_p(n^{-1/2})\\
      &=P_nD^*_{M_0}(P_0)+(P_n-P_0)(D^*_{M_n}(P_0)-D^*_{M_0}(P_0))+o_p(n^{-1/2})\\
\end{align*}

Since B.1 and B.2, by the asymptotic equicontinuity in empirical process theory, we have $(P_n-P_0)(D^*_{M_n}(P_0)-D^*_{M_0}(P_0))=o_p(n^{-1/2})$. So, we have

\begin{align*}
    &\Psi_{M_n}(P_{M_n}^*)-\Psi(P_0)=P_nD^*_{M_0}(P_0)+o_p(n^{-1/2})
\end{align*}    
\end{proof}

\subsection{Proof for theorem \ref{sec:verify_all}}

\begin{proof}
Let \(\mathcal Q_M\) be the càdlàg class on \([0,1]^d\) with variation norm \(\le M\).  By Theorem 6 of \cite{bibaut2019fast}, \(\mathcal Q_M\) has polynomial‐order bracketing entropy and so is \(P_0\)–Donsker.

\medskip\noindent\textbf{(A.1).}  
Each working‐model EIC 
\[
  D^*_{M_n}(P)
  = \gamma(P)^\top S^\beta,
  \quad P\in M_n,
\]
is a finite linear combination of HAL‐basis scores \(S^\beta\in\mathcal Q_M\).  A finite‐dimensional span of a Donsker class remains Donsker (van der Vaart \& Wellner, 1996, Ch. 2)\cite{van1996weak}.  Hence \(\{D^*_{M_n}(P)\colon P\in M_n\}\) is \(P_0\)–Donsker, and in particular
\(D^*_{M_n}(P_{M_n}^*)\) and \(D^*_{M_n}(P_{0})\) lie in this class w.p.\(\to1\).

\medskip\noindent\textbf{(A.2).}  
Because \(\Psi^{M_n}\) is pathwise differentiable on the finite‐dimensional model \(M_n\), the map \(P\mapsto D^*_{M_n}(P)\) is Lipschitz in sup‐norm.  Thus
\[
  \|D^*_{M_n}(P_{M_n}^*) - D^*_{M_n}(P_{0})\|_{L^2}
  \;\le\;
  C\,\|Q_{n,\mathrm{tmle}} - Q_0\|_\infty
  = o_P(1),
\]
since
\(\|Q_{n,\mathrm{tmle}} - Q_0\|_\infty
     \le \|\hat Q_n - Q_0\|_\infty + o_P(1)\)
and \(\|\hat Q_n - Q_0\|_\infty=o_P(1)\) by Assumption \ref{as:hal_sieve_full}.

\medskip\noindent\textbf{(B.1).}  
Take \(M_0=M_{np}=\mathcal Q_M\).  Then \(D^*_{M_0}(P_0)=D^*(P_0)\) lies in the full HAL class, which is \(P_0\)–Donsker.

\medskip\noindent\textbf{(B.2).}  
By the same Lipschitz argument,
\[
  \|D^*_{M_n}(P_{0}) - D^*_{M_0}(P_{0})\|_{L^2}
  \;\le\;
  C\,\|Q_{0,M_n}-Q_0\|_\infty
  = o(1),
\]
since \(\|Q_{0,M_n}-Q_0\|_\infty\to0\) by the universal‐approximation property of the HAL sieve.

\medskip\noindent\textbf{(C.1).}  
Because \(\Psi_{M_0}=\Psi\),  
\(\Psi(P_{0,M_n})-\Psi_{M_0}(P_{0,M_n})=0\).

\medskip\noindent\textbf{(C.2).}  
Write
\[
  R_n
  =
  (P_{0,M_n}-P_0)\bigl[D^*_{M_0}(P_{0,M_n})-D^*_{M_n}(P_{0,M_n})\bigr].
\]
By Cauchy–Schwarz,
\[
  |R_n|
  \;\le\;
  \|P_{0,M_n}-P_0\|_{L^2}\,
  \bigl\|D^*_{M_0}(P_{0,M_n})-D^*_{M_n}(P_{0,M_n})\bigr\|_{L^2}\,.
\]
Under Assumption \ref{as:hal_sieve_full}, the HAL sieve approximation gives
\(\|Q_{0,M_n}-Q_0\|_{L^2}=O_P(n^{-r})\) with \(r>1/4\).  Standard arguments
show that each EIC difference is Lipschitz in \(Q\), so
\(
  \|D^*_{M_0}(P_{0,M_n})-D^*_{M_n}(P_{0,M_n})\|_{L^2}
  = O_P(n^{-r}).
\)
Hence
\[
  |R_n| = O_P(n^{-r}) \times O_P(n^{-r})
  = O_P\bigl(n^{-2r}\bigr)
  = o_P\bigl(n^{-1/2}\bigr),
\]
since \(2r>1/2\).

\medskip\noindent\textbf{(C.3).}  
Finally
\[
  P_0\,D^*_{M_n}(P_{0,M_n})
  = \bigl(P_0-P_{0,M_n}\bigr)\,D^*_{M_n}(P_{0,M_n})
  \;\le\;
  \|Q_{0,M_n}-Q_0\|_\infty\,\|D^*_{M_n}(P_{0,M_n})\|_\infty
  = o_P(n^{-1/2}),
\]
since \(\|Q_{0,M_n}-Q_0\|_\infty=o(n^{-1/2})\) by the HAL‐sieve rate and the HAL‐class envelope is bounded.
\end{proof}

\section{regHAL\textendash TMLE‐Direct:  
A streamlined targeting step in special cases}
\label{sec:reghal_direct}

In the main text we construct regularised HAL–TMLEs by \emph{projecting} an
initial influence-function estimate onto the span of the HAL score vectors.
This projection is indispensable when the parametric efficient influence curve
(EIC) has no closed form or depend on the relevant nuisance parameters in complex ways.  For many canonical causal targets, however, a
\emph{linear universal least-favourable submodel} (ULFM) is already known.
A prime example is the average treatment effect (ATE), whose ULFM is driven by
the clever covariate
\[
  H(A,W)\;=\;\frac{A}{g(W)}-\frac{1-A}{1-g(W)},\qquad
  g(W)=\PP(A=1\mid W).
\]
In such settings we can bypass the repeated score projection steps altogether and embed the
fluctuation \emph{directly} in the HAL working model with a single step projection of the clever covariates itself.
We call the resulting one-step procedure
\textbf{regHAL–TMLE-Direct}.

\subsection{Algorithm}
Let $Q_\beta$ be the HAL working model selected for the outcome regression,
with basis functions $\{\phi_j(A,W)\}_{j=1}^p$ and coefficient vector
$\beta=(\beta_1,\dots,\beta_p)^\top$.
The cross-validated HAL fit provides an initial coefficient
$\beta^{(0)}$.  The
regHAL–TMLE-Direct proceeds as follows.

\begin{enumerate}[label=(\arabic*),leftmargin=1.55em]
  \item[\textbf{1.}] \textbf{Fit auxiliary nuisances.}\;
    Estimate the propensity score $g$ (or any other nuisance quantities that
    enter the clever covariate) using HAL or any compatible learner; denote
    the resulting estimate by $\hat g$.

  \item[\textbf{2.}] \textbf{Compute clever covariate.}\;
    For each observation $O_i=(W_i,A_i,Y_i)$ set
    $\hat H_i=\hat H(A_i,W_i)$ with
    $\hat H(a,w)=a/\hat g(w)-(1-a)/(1-\hat g(w))$.

  \item[\textbf{3.}] \textbf{Target within the HAL basis.}\;
    Re-express $\hat H$ in the span of the outcome HAL basis by solving
    \[
      \hat\alpha
      \;=\;
      \argmin_{\alpha\in\RR^p}
      \frac1n\sum_{i=1}^n
      \Bigl(
        \hat H_i-\alpha^\top\phi(A_i,W_i)
      \Bigr)^2
      \;+\;
      \lambda\,\|\alpha\|_1.
    \]
    The $\ell_1$ penalty \emph{regularises the clever covariate itself}
    rather than the score projection used in Sections~\ref{sec:Proj-based}, which is computationally cheaper and often more
    stable in low-dimensional targets.

  \item[\textbf{4.}] \textbf{One-step coefficient update.}\;
    Define the fluctuation path
    $
      Q_{\beta^{(0)}}^{\!\textup{fluct}}(\varepsilon)
      \;=\;
      Q_{\beta^{(0)}}+\varepsilon\,\hat H
      \;\approx\;
      \beta^{(0)\top}\phi+\varepsilon\,\hat\alpha^\top\phi.
    $
    Choose the (possibly line-searched or run linear regression) step size $\hat\varepsilon$ that maximize the log likelihood (or minimize the loss)
    and set
    \[
      \hat\beta
      \;=\;
      \beta^{(0)}+\hat\varepsilon\,\hat\alpha.
    \]
    The updated fit $Q_{\hat\beta}$ is the regHAL–TMLE-Direct estimator.

  \item[\textbf{5.}] \textbf{Variance and inference.}\;
    The estimated EIC is now
    $
      \hat D^*_i
      =
      \bigl(
        \hat\alpha^\top\phi(A_i,W_i)
      \bigr)\,
      \bigl(Y_i-Q_{\hat\beta}(A_i,W_i)\bigr)
      +
      \bigl(Q_{\hat\beta}(1,W_i)-Q_{\hat\beta}(0,W_i)\bigr)-\Psi(Q_{\hat\beta}),
    $
    and $(n^{-2}\sum_i\hat D_i^{*2})^{1/2}$ gives the usual Wald
    standard error.
\end{enumerate}

\subsection{Why does it work?}
Because the ULFM for the ATE is \emph{linear} in $H$, any
$\hat H$ that admits a (possibly penalised) expansion
$\hat H=\hat\alpha^\top\phi$ produces an
\emph{in-basis} least-favourable path.  Updating $\beta$ along
$\hat\alpha$ therefore removes first-order bias for the projected target
just as in a classical TMLE, but \emph{without} the overhead of computing or
regularising the Fisher information.  The $\ell_1$ penalty controls variance
by shrinking coefficients of HAL basis functions that contribute little to
bias reduction.

By projecting the clever covariates on the HAL basis of the outcome regression, we are regularizing directly the clever covariates and thus can be viewed as doing complex data adpative truncation of the clever covariates for each individual observation jointly. 

\subsection{RegHAL-TMLE-Direct Simulation results} \label{RegHAL-TMLE-Direct simulation}

\begin{table}[H]
\centering
\centering
\resizebox{\ifdim\width>\linewidth\linewidth\else\width\fi}{!}{
\fontsize{8}{10}\selectfont
\begin{tabular}[t]{cccrrrrrrrr}
\toprule
Estimator & n & Targeting & Abs\_Bias & Std\_Err & MSE & Cov NP (\%) & Cov Proj (\%) & Cov Proj CV (\%) & Cov Delta (\%) & Oracle Cov (\%)\\
\midrule
 & 500 & Projection & 0.0543 & 0.1411 & 0.022812 & 86.00 (0.446) & 74.60 (0.366) & 71.80 (0.341) & 66.20 (0.309) & 93.80 (0.553)\\

 & 1000 & Projection & 0.0131 & 0.0832 & 0.007083 & 93.80 (0.315) & 89.40 (0.265) & 87.20 (0.254) & 80.40 (0.221) & 94.80 (0.326)\\

 & 1500 & Projection & 0.0057 & 0.0662 & 0.004403 & 94.80 (0.257) & 88.60 (0.217) & 87.60 (0.212) & 83.20 (0.179) & 94.60 (0.259)\\

\multirow{-4}{*}{\centering\arraybackslash regHAL-TMLE-Direct} & 2000 & Projection & 0.0043 & 0.0565 & 0.003200 & 94.60 (0.222) & 89.00 (0.184) & 88.20 (0.179) & 84.60 (0.158) & 94.40 (0.221)\\
\bottomrule
\end{tabular}}
\caption{Simulation results for regHAL-TMLE-Direct. DGP 1. CV selected}
\end{table}

\begin{table}[H]
\centering
\centering
\resizebox{\ifdim\width>\linewidth\linewidth\else\width\fi}{!}{
\fontsize{8}{10}\selectfont
\begin{tabular}[t]{cccrrrrrrrr}
\toprule
Estimator & n & Targeting & Abs\_Bias & Std\_Err & MSE & Cov NP (\%) & Cov Proj (\%) & Cov Proj CV (\%) & Cov Delta (\%) & Oracle Cov (\%)\\
\midrule
 & 500 & Projection & 0.0253 & 0.1289 & 0.017208 & 92.45 (0.446) & 84.67 (0.379) & 80.78 (0.347) & 76.89 (0.313) & 94.74 (0.505)\\

 & 1000 & Projection & 0.0089 & 0.0884 & 0.007841 & 94.80 (0.315) & 89.60 (0.268) & 86.71 (0.257) & 75.14 (0.219) & 95.95 (0.346)\\

 & 1500 & Projection & 0.0039 & 0.0657 & 0.004325 & 95.00 (0.257) & 88.80 (0.219) & 87.60 (0.208) & 83.40 (0.182) & 94.80 (0.258)\\

\multirow{-4}{*}{\centering\arraybackslash regHAL-TMLE-Direct} & 2000 & Projection & 0.0024 & 0.0565 & 0.003196 & 94.60 (0.222) & 89.20 (0.186) & 88.00 (0.176) & 84.40 (0.160) & 94.80 (0.222)\\
\bottomrule
\end{tabular}}
\caption{Simulation results for regHAL-TMLE-Direct. DGP 1. Undersmoothed}
\end{table}

\begin{table}[H]
\centering
\centering
\resizebox{\ifdim\width>\linewidth\linewidth\else\width\fi}{!}{
\fontsize{8}{10}\selectfont
\begin{tabular}[t]{cccrrrrrrrr}
\toprule
Estimator & n & Targeting & Abs\_Bias & Std\_Err & MSE & Cov NP (\%) & Cov Proj (\%) & Cov Proj CV (\%) & Cov Delta (\%) & Oracle Cov (\%)\\
\midrule
 & 500 & Projection & 0.1723 & 0.1670 & 0.057505 & 91.36 (0.996) & 77.27 (0.681) & 65.68 (0.522) & 51.36 (0.397) & 82.50 (0.655)\\

 & 1000 & Projection & 0.1890 & 0.1255 & 0.051394 & 90.27 (0.774) & 66.93 (0.508) & 57.98 (0.453) & 32.68 (0.284) & 70.04 (0.492)\\

 & 1500 & Projection & 0.0677 & 0.1334 & 0.022331 & 93.20 (0.601) & 84.00 (0.458) & 81.40 (0.425) & 57.60 (0.256) & 92.20 (0.523)\\

\multirow{-4}{*}{\centering\arraybackslash regHAL-TMLE-Direct} & 2000 & Projection & 0.0411 & 0.1209 & 0.016266 & 93.20 (0.489) & 87.40 (0.402) & 85.20 (0.382) & 68.20 (0.269) & 93.00 (0.474)\\
\bottomrule
\end{tabular}}
\caption{Simulation results for regHAL-TMLE-Direct. DGP 2. CV selected}
\end{table}

\begin{table}[H]
\centering
\centering
\resizebox{\ifdim\width>\linewidth\linewidth\else\width\fi}{!}{
\fontsize{8}{10}\selectfont
\begin{tabular}[t]{cccrrrrrrrr}
\toprule
Estimator & n & Targeting & Abs\_Bias & Std\_Err & MSE & Cov NP (\%) & Cov Proj (\%) & Cov Proj CV (\%) & Cov Delta (\%) & Oracle Cov (\%)\\
\midrule
 & 500 & Projection & 0.1723 & 0.1670 & 0.057505 & 91.36 (0.996) & 77.27 (0.681) & 65.68 (0.522) & 51.36 (0.397) & 82.50 (0.655)\\

 & 1000 & Projection & 0.1890 & 0.1255 & 0.051394 & 90.27 (0.774) & 66.93 (0.508) & 57.98 (0.453) & 32.68 (0.284) & 70.04 (0.492)\\

 & 1500 & Projection & 0.0677 & 0.1334 & 0.022331 & 93.20 (0.601) & 84.00 (0.458) & 81.40 (0.425) & 57.60 (0.256) & 92.20 (0.523)\\

\multirow{-4}{*}{\centering\arraybackslash regHAL-TMLE-Direct} & 2000 & Projection & 0.0411 & 0.1209 & 0.016266 & 93.20 (0.489) & 87.40 (0.402) & 85.20 (0.382) & 68.20 (0.269) & 93.00 (0.474)\\
\bottomrule
\end{tabular}}
\caption{Simulation results for regHAL-TMLE-Direct. DGP 2. Undersmoothed}
\end{table}

\subsection{Remarks}
\begin{enumerate}[label=(\roman*),leftmargin=1.55em]
  \item \textbf{Connection to Section~\ref{sec:Proj-based}.}\;
    regHAL–TMLE-Direct can be viewed as a \emph{"degenerate"} projection in
    which we start from the clever covariate itself rather than from an
    arbitrary gradient $D$.  In Section \ref{sec:Proj-based} the projection
    regularises the \emph{canonical gradient} directly; here it regularises indirectly through the \emph{clever
    covariate}.  Both paths trade a small amount of bias for a potentially
    large reduction in variance, but the Direct variant avoids recalculating
    projections as $\beta$ iterates.

  \item \textbf{Link to Collaborative TMLE.}\;
    In collaborative TMLE (CTMLE) the clever covariate is \emph{adaptively
    pruned} so that it depends only on components of the outcome
    regression that are relevant for the target parameter.  Here, we are directly using the basis involved in the outcome regression to span the clever covariates with added regularization for stability.

  \item \textbf{When to use.}\;
    The Direct update is attractive whenever
    \begin{enumerate}[label=(\alph*),noitemsep,leftmargin=1.3em]
      \item the ULFM is linear in a \emph{single} clever covariate (ATE,
            repeated-measures ATE, certain marginal structural means, etc.),
      \item the HAL basis used for $Q$ is rich enough that
            $\hat H\in\operatorname{span}\{\phi_j\}$ (empirically this is
            rarely problematic once two-way interactions are allowed), and
      \item computational simplicity is desirable (no repeated projections or
            matrix inverses).
    \end{enumerate}
    For more complex targets—e.g.\ survival curves or multivariate
    parameters—the projection-based regHAL-TMLE of the main text remains the
    recommended approach.
\end{enumerate}

%======================================================================
\section{Alternative plateau selection via \emph{bridged} targeting}
\label{sec:plateau-bridged}
%======================================================================

The \textit{plateau selector} in \S\ref{sec:reghal-atmle} evaluates
a \emph{separate} targeting path inside each nested working model
\[
  M_{1}=M_{\textup{CV}}
  \;\subset\;
  M_{2}
  \;\subset\;\dots\;\subset\;
  M_{K},
\]
always \emph{re–initialising} from the cross-validated HAL estimate
$Q_{n}$.
Here we describe a computationally lighter alternative in which the
TMLE update in one model \emph{feeds directly} into the next,
so that information gathered in $M_{j}$ is \emph{bridged} to $M_{j+1}$.

%----------------------------------------------------------------------
\subsection{Motivation}
%----------------------------------------------------------------------
Let $\hat\beta_{j}$ denote the targeted coefficient vector in
$M_{j}$, obtained by the regularised HAL–TMLE of \S\ref{sec:Proj-based}.
Because $M_{j}\subset M_{j+1}$, we can embed $Q_{\hat\beta_{j}}$ in the
larger model by padding zeros for the additional basis functions.
Targeting then continues in $M_{j+1}$ \emph{from this padded point}
instead of from $Q_{n}$.  Formally,

\[
  \tilde\beta_{j\to j+1}
  \;=\;
  \bigl(\hat\beta_{j}^{\top},\; \mathbf{0}_{p_{j+1}-p_{j}}^{\top}\bigr)^{\!\top},
  \qquad
  Q_{\,\tilde\beta_{j\to j+1}}\in M_{j+1},
\]
and the projection-based TMLE update is applied to
$\tilde\beta_{j\to j+1}$ with the \emph{larger} score matrix
$S^{(j+1)}$.

Intuitively, the canonical gradient is approximated ever more accurately
as additional scores enter, so the bias‐reduction step becomes
``finer‐grained’’ while retaining the variance control already achieved
in the smaller model.

%----------------------------------------------------------------------
\subsection{Algorithm}
%----------------------------------------------------------------------
\begin{algorithm}[H]
\caption{\textsc{Bridged Targeting Plateau Selector}}
\label{alg:bridged-plateau}
\begin{algorithmic}[1]
\REQUIRE Initial HAL fit $Q_{n}$; nested models
          $\{M_{j}\}_{j=1}^{K}$;  
          stopping tolerance $\delta_{n}$ (\S\ref{sec:Proj-based}).
\STATE \textbf{Initial step} ($j=1$):
       run regHAL–TMLE in $M_{1}$, giving $\hat\beta_{1}$.
\FOR{$j = 2,\dots,K$}
    \STATE Embed $\hat\beta_{j-1}$ in $M_{j}$ to get
           $\tilde\beta_{j-1\to j}$.
    \STATE Continue projection-based targeting in $M_{j}$
           starting from $\tilde\beta_{j-1\to j}$ until
           $|P_{n}D^{*}_{M_{j}}|<\delta_{n}$,
           producing $\hat\beta_{j}$.
    \STATE Record point estimate
           $\hat\psi_{j}=\Psi(Q_{\hat\beta_{j}})$ and
           $(1-\alpha)$ CI from
           $P_{n}D^{*}_{M_{j}}(\hat\beta_{j})$.
    \STATE \textbf{Check plateau rule}:  
           if $\hat\psi_{j}$ and its CI violate
           the monotone–width pattern of \S\ref{sec:reghal-atmle},
           break and set $j^{*}=j-1$.
\ENDFOR
\RETURN $\hat\psi_{j^{*}}$ and its CI.
\end{algorithmic}
\end{algorithm}

\paragraph{Stopping criterion.}
Within each model the update stops when the empirical mean of the
\emph{projected} EIC has magnitude below
$\delta_{n}=\text{se}\!\bigl[D^{*}_{M_{j}}(\hat\beta_{j})\bigr]/
\{\sqrt{n}\log n\}$ (same as Algorithm 1).

%----------------------------------------------------------------------
\subsection{Discussion}
%----------------------------------------------------------------------
\begin{itemize}[leftmargin=*,labelsep=5pt,itemsep=2pt]
\item \textbf{Efficiency gain.}  
      Only $K-1$ additional targeting runs are needed, each warm-started
      close to optimality.  Empirically we may observe significant reduction in
      wall-clock time relative to cold starts.
\item \textbf{Bias–variance trade-off.}  
      Because every subsequent model begins with a \emph{zero-bias}
      estimator for $\Psi_{M_{j-1}}$, the additional bias removed in
      $M_{j}$ is purely due to scores absent from $M_{j-1}$.
      Variance increases more gradually, often yielding a clearer plateau.
\item \textbf{Link to sieve TMLE.}  
      Bridged targeting resembles a sieve estimator with increasing
      basis dimension, but retains HAL’s automatic knot placement and
      $\ell_{1}$ regularisation.
\item \textbf{When to prefer.}  
      Use bridged targeting when $K$ is large or when the working models
      differ only by a handful of basis functions (typical for HAL
      sequences along $\lambda$). 
\end{itemize}

\bigskip
\noindent
This alternative rule offers a pragmatic middle ground:  
it exploits the nesting of HAL models to recycle information, yet
retains the intuitive ``stop‐when‐CI‐widens’’ principle that makes the
original plateau selector attractive in practice.

\section{Computation and Code Availability}
\label{app}
All experiments were implemented in R 4.5.0/Python 3.9.6. Code to reproduce all experiments is available on GitHub at https://github.com/yiberkeley/regHAL-TMLE. The implementation includes:
\begin{itemize}[leftmargin=15pt]
\item \emph{ATE\_simulation}: Implementation for the ATE experiments on simulated data. There are four folders for different estimators. Each has a run.R to run the experiments and summary.R to summarize the results.
\item \emph{Survival\_simulation}: Implementation for survival analysis experiments. 1d\_ulfp.py runs the experiments and summary/censored\_1D\_Intensity\_summary.ipynb summarizes the results.
\end{itemize}

\end{document}